\documentclass[preprint,3p,10pt]{elsarticle}
\usepackage{srcltx}
\usepackage{times,amsbsy,amsmath,amsthm,amssymb,amscd,amsfonts,mathrsfs}
\usepackage{graphicx,epstopdf}
\usepackage[hang]{subfigure}
\usepackage{threeparttable}
\usepackage{multirow}
\usepackage{dcolumn}
\usepackage{booktabs}
\usepackage{tikz}
\usetikzlibrary{arrows, snakes, backgrounds}
\usepackage{float}
\usepackage{color}
\usepackage{algorithm}
\usepackage{algorithmic}

\usepackage{lineno,hyperref}
\modulolinenumbers[5]

\journal{Journal of \LaTeX\ Templates}

\begin{document}
\begin{frontmatter}

\title{A conservative discrete velocity method for the ellipsoidal Fokker-Planck equation in gas-kinetic theory}
\author[authorlabel1,authorlabel2]{Sha liu\corref{cor1}}
\ead{shaliu@nwpu.edu.cn}
\cortext[cor1]{Corresponding author}
\author[authorlabel2]{Ruifeng Yuan}
\ead{xyrfx@mail.nwpu.edu.cn}
\author[authorlabel2]{Usman Javid}
\ead{NormiJavid27@mail.nwpu.edu.cn}
\author[authorlabel1,authorlabel2]{Chengwen Zhong}
\ead{zhongcw@nwpu.edu.cn}

\address[authorlabel1]{National Key Laboratory of Science and Technology on Aerodynamic Design and Research, Northwestern Polytechnical University, Xi'an, Shaanxi 710072, China}
\address[authorlabel2]{School of Aeronautics, Northwestern Polytechnical University, Xi'an, Shaanxi 710072, China}

\begin{abstract}
A conservative discrete velocity method (DVM) is developed for the ellipsoidal Fokker-Planck (ES-FP) equation in prediction of non-equilibrium neutral gas flows in this paper. The ES-FP collision operator is solved in discrete velocity space in a concise and quick finite difference framework. The conservation problem of discrete ES-FP collision operator is solved by multiplying each term in it by extra conservative coefficients whose values are very closed to unity. Their differences to unity are in the same order of the numerical error in approximating the ES-FP operator in discrete velocity space. All the macroscopic conservative variables (mass, momentum and energy) are conserved in the present modified discrete ES-FP collision operator. Since the conservation property in discrete element of physical space is very important for numerical scheme when discontinuity and large gradient exist in flow field, a finite volume framework is adopted for the transport term of ES-FP equation. For $nD$-$3V$ ($n<3$) cases, a $nD$-quasi $nV$ reduction is specially proposed for ES-FP equation and the corresponding FP-DVM method, which can greatly reduce the computational cost. The validity and accuracy of both ES-FP equation and FP-DVM method are examined using a series of $0D$-$3V$ homogenous relaxation cases and $1D$-$3V$ shock structure cases with different $Mach$ numbers, in which $1D$-$3V$ cases are reduced to $1D$-quasi $1V$ cases. Both the predictions of $0D$-$3V$ and $1D$-$3V$ cases match well with the benchmark results such as analytical Boltzmann solution, direct full-Boltzmann numerical solution and DSMC result. Especially, the FP-DVM predictions match well with the DSMC results in the $Mach$ 8.0 shock structure case, which is in high non-equilibrium, and is a challenge case of the model Boltzmann equation and the corresponding numerical methods.
\end{abstract}

\end{frontmatter}

\section{Introduction}
The Fokker-Planck (FP) equation with advection-diffusion collision operator is widely used in modeling dynamic systems such as neutral molecule~\cite{K2017parallel, Gorji2015Fokker, Mathiaud2016A}, plasma~\cite{rosenbluth1957fokker, landau1958kinetic, Degond1994An}, photonics~\cite{kompaneets1957establishment, zel1972stimulated}, and even biological~\cite{zeng2018distribution}, economic~\cite{dolfin2017modeling}, and social~\cite{toscani2018opinion} systems. The first FP equation for molecule system is derived from Boltzmann equation in gas kinetic theory when counting the gazing effect of molecule collisions~\cite{Cercignani1990The}. The Prandtl ($Pr$) number yielded from this FP equation is fixed at $3/2$. Since the $Pr$ number of real gas is below unity ($2/3$ for monatomic gas), two types of modified FP equations are proposed. They are cubic-FP equation~\cite{Gorji2015Fokker} and ellipsoidal-FP (ES-FP) equation~\cite{Mathiaud2016A}. Recently, by mapping these FP equations to Stochastic Differential Equations (SDEs), the FP equations are solved in a stochastic and particle way~\cite{K2017parallel, Gorji2015Fokker}. Comparing with other particle methods such as Direct Simulation Monte-Carlo (DSMC)~\cite{bird2013dsmc}, its computational cost is greatly reduced in continuum limit (dissipation limit in FP research). Since the mechanism of drag and diffusion forces holds for both micro and macro scales, then large time step and cell length can be used. This advantage is very important for the prediction of flow fields in either transitional or continuum regimes where the molecular mean free path (m.f.p.) and mean collision time (m.c.t.) are comparable or greatly less than the characteristic length and time, respectively. On the other hand, since the deterministic methods are not affected by statistical fluctuation, they are very useful in the precise computation of multi-scale non-equilibrium flows, and are helpful in investigating the mechanism of such flows. Moreover, the accuracy of modified FP equations, especially their collision operators, should be examined using a deterministic method.

The first deterministic numerical scheme for FP equation, which can preserve equilibrium, is proposed for a $1V$ (one dimensional velocity space) isotropic linear Fokker-Planck-Landau (FPL) system~\cite{Chang1970A}. Then it is extended to mass/energy/equilibrium preserving scheme~\cite{Buet2007Positive}, $2V$ cases in discrete cylindrical velocity space~\cite{Yoon2014Erratum}, and nonlinear equation~\cite{Larsen1985Discretization, Epperlein1994Implicit}. Besides the FPL-type, another form of FP equation is Rosenbluth-Fokker-Planck equation (RFP)~\cite{rosenbluth1957fokker}. RFP equation has a similar mathematical form of the FP equation that is derived from Boltzmann equation~\cite{Cercignani1990The}. Its collision operator is written in a differential form with nonlinear advection and diffusion coefficients. By using a finite volume framework in velocity space and extra coefficients on advection terms for conservation purpose, RFP equation is well solved by the deterministic numerical method in Ref.~\cite{Taitano2015A, Taitano2017An}.

Plenty of the FP researches focus on the homogenous FP equation where the particle transport in physical space is assumed to be zero~\cite{Crouseilles2004Numerical}. Several works address the non-homogenous FP equation in the topic of FPL-type equation~\cite{filbet2002numerical, duclous2009high}. In flow predictions, the particle transport in physical space is an essential aspect that can not be ignored. The transport term in the FP equation of gas-kinetic theory (similar to RFP-type) should be calculated. By taking the particle transport into consideration, the deterministic numerical framework will be the discrete velocity method (DVM). There are several multi-scale methods in gas kinetic theory, such as Unified Gas-Kinetic Scheme (UGKS)~\cite{xu2010unified}, Discrete Unified Gas-Kinetic Scheme (DUGKS)~\cite{guo2013discrete}, Gas-Kinetic Unified Algorithm (GKUA)~\cite{peng2016implicit}, using the DVM framework for predictions of flows from continuum regime to rarefied regime. Recently, these methods have been widely used in the prediction of non-equilibrium flows~\cite{xu2010unified}, plasmas~\cite{Liu2017A}, and photonics~\cite{Guo2016Discrete}. Since they use Bhatnagar-Gross-Krook (BGK)-type model equations, certain degree of deviation (from the Boltzmann equation) exists in the prediction of high non-equilibrium flows~\cite{Liu2014Investigation}, such as the shock structure cases with high $Mach$ numbers, which will be calculated in the later section.

In this paper, a novel deterministic method (FP-DVM) is proposed for non-equilibrium flows, which solves the ES-FP equation numerically in the DVM framework. In FP-DVM, the ES-FP collision operator is solved in a deterministic way in discrete velocity space using a framework of Finite Difference Method (FDM). Instead of considering the conservation in discrete element of velocity space, the discrete ES-FP collision operator is treated in a relaxation way, and the conservations of mass, momentum and energy are ensured by coefficients being added to both advection and dissipation terms. The deviations of their values from unity have the same order as the truncation error of the numerical approximation to ES-FP collision operator. The computational complexity of solving the discrete ES-FP collision operator is $O(N)$, here ``$N$" is the number of discrete points in velocity space. Four $0D$-$3V$ ($nD$-$mV$ is a denotation of cases whose dimension of physical space is ``$n$" and the dimension of velocity space is ``$m$" in the scope of FP research) homogenous relaxation cases and three $1D$-$3V$ shock structure cases with different $Mach$ numbers are conducted to examine the validity and accuracy of both ES-FP equation and the present numerical method. Using the dimensional reduction method for ES-FP equation proposed in this paper, $1D$-$3V$ cases are reduced to $1D$-quasi $1V$ cases, and the computational cost is greatly reduced. The remaining of this paper is organized as follows: Sec.~\ref{sec:2} is a quick review of gas-kinetic theory and FP equation. Sec.~\ref{sec:3} is the construction of FP-DVM method; Sec.~\ref{sec:4} is the numerical experiment; Sec.~\ref{sec:5} is the discussion and conclusion.

\section{Gas kinetic theory and Fokker-Planck equation}\label{sec:2}
\subsection{Distribution function and Boltzmann equation}
In gas kinetic theory, molecular system is described using distribution function $f(x_i,\xi_j,t)$ depending on location $x_i$, molecular velocity $\xi_j$ and time $t$. It is the number density of molecules that arrived at $x_i$ at time $t$ with velocity $\xi_j$. For dilute gas, the evolution of $f$ is governed by Boltzmann equation~\cite{Kremer2010An}:
\begin{equation}\label{eq:Boltzmann}
\frac{{\partial f}}{{\partial t}} + {\xi _i}\frac{{\partial f}}{{\partial {x_i}}} + {a_i}\frac{{\partial f}}{{\partial {\xi _i}}} = C\left( {f,f} \right),
\end{equation}
where $a_i$ is acceleration of molecule due to body force such as gravity. Einstein summation convention is used throughout this paper if without special statement. The Left Hand Side (LHS) of Eq.~\ref{eq:Boltzmann} is the free transport operator, while the Right Hand Side (RHS) is the collision operator which is mathematically a five-fold nonlinear integral.

Given the distribution function $f$, macroscopic physical variables, such as mass density $\rho$, momentum density $\rho u_i$ ($u_i$ is macroscopic velocity), energy density $\rho e$ ($e$ is energy per mass), stress $\tau_{ij}$ and heat flux $q_i$, can be obtained using their definition in gas-kinetic theory as follows,
\begin{equation}\label{eq:constrain}
\begin{aligned}
&\rho  = \left\langle {mf} \right\rangle = mn, \\
&\rho {u_i} = \left\langle {m{\xi _i}f} \right\rangle, \\
&\rho e = \left\langle {\frac{1}{2}m{\xi _k}{\xi _k}f} \right\rangle = \frac{1}{2}\rho u_{k}u_{k} + \frac{3}{2}nkT, \\
&{\tau _{ij}} = -\left\langle {m\left({c_i}{c_j}-\frac{1}{3}{c_k}{c_k}\delta_{ij}\right)f}\right\rangle = -nk\left(T_{ij} - T\delta_{ij}\right),\\
&{q_i} = \left\langle {\frac{1}{2}m{c_i}{c_k}{c_k}} \right\rangle,
\end{aligned}
\end{equation}
where $n$ is number density, $c_i$ is the peculiar velocity defined as $\xi_i-u_i$, $T$ is thermodynamic temperature, $T_{ij}$ is the temperature tensor in gas kinetic theory whose trace is $3T$, $k$ is Boltzmann constant, $m$ is the mass of molecule, $\delta_{ij}$ is the Kronecker delta, the operator $\left\langle  \cdot  \right\rangle$ is an integral over the whole velocity space, which can be written as
\begin{equation}
\left\langle  \cdot  \right\rangle = \int_{ - \infty }^{ + \infty } {\int_{ - \infty }^{ + \infty } {\int_{ - \infty }^{ + \infty } {\left(  \cdot  \right)d{\xi _{1}}d{\xi _{2}}d{\xi _{3}}} } }.
\end{equation}

\subsection{Fokker-Planck equation for gas kinetic theory}
In the scope of gas kinetic theory, the original Fokker-Planck equation without body force is derived in Ref.~\cite{Cercignani1990The}, where the grazing effect of binary collisions is considered. By changing its relaxation rate to $\tau_{FP}=2\mu/p$ ($\mu$ is viscosity, $p$ is pressure), the standard Fokker-Planck equation as a model of the Boltzmann equation can be written as follows,
\begin{equation}\label{eq:FP}
\frac{{\partial f}}{{\partial t}} + {\xi _i}\frac{{\partial f}}{{\partial {x_i}}} = \frac{1}{{{\tau _{FP}}}}\left\{ {\frac{{\partial \left( {\left( {{\xi _i} - {u_i}} \right)f} \right)}}{{\partial {\xi _i}}} + RT{\delta _{ij}}\frac{{{\partial ^2}f}}{{\partial {\xi _i}\partial {\xi _j}}}} \right\},
\end{equation}
where $R=k/m$ is the specific gas constant. Since the standard Fokker-Planck equation corresponds to a fix Prandtl number of $3/2$, two types of modified Fokker-Planck equations, the cubic-FP equation equation~\cite{Gorji2015Fokker} and ES-FP equation~\cite{Mathiaud2016A}, are proposed. In cubic-FP, the advection term is multiplied by a polynomial of molecular velocity $\xi_{i}$, whose coefficients are used to get the right relaxation rate of both stress and heat flux, thus a right Prandtl is realized. In ES-FP equation, the diagonal dissipation coefficient $RT\delta_{ij}$ in the standard FP equation is replaced by $T_{ES,ij}$ which is defined as follows
\begin{equation}
{T_{ES,ij}} = \left( {1 - \nu } \right)T{\delta _{ij}} + \nu {T_{ij}},
\end{equation}
as a combination of isotropic temperature $T$ and anisotropic temperature $T_{ij}$ ($T_{ij}=\left\langle{m{c_i}{c_j}f}\right\rangle/\rho R$), and $\nu$ is defined as
\begin{equation}\label{eq:nu}
\nu  = \max \left( { - \frac{5}{4}, - \frac{{T}}{{{\lambda _{\max }} - T}}} \right),
\end{equation}
where $\lambda_{max}$ is the maximum eigenvalue of the positive definite matrix $T_{ij}$. Since $\nu$ and $Pr$ number have the following relation
\begin{equation}
\Pr  = \frac{3}{{2\left( {1 - \nu } \right)}},
\end{equation}
the $Pr$ number is $2/3$ except in the extreme condition $\lambda_{\max}>1.8T$. In this extreme condition, $Pr$ varies from $2/3$ to unity. The relaxation time $\tau_{ES}$ in ES-FP collision operator is defined as
\begin{equation}
{\tau _{ES}} = 2\left( {1 - \nu } \right)\frac{\mu }{p}.
\end{equation}
Since the procedure of deriving ES-FP equation from standard FP equation is similar to that of extending BGK equation to ES-BGK equation, it is called ES-FP in Ref.~\cite{Mathiaud2016A}. Similar to standard FP, ES-FP is written in the following form
\begin{equation}\label{eq:ES-FP}
\frac{{\partial f}}{{\partial t}} + {\xi _i}\frac{{\partial f}}{{\partial {x_i}}} = \frac{1}{{{\tau _{ES}}}}\left\{ {\frac{{\partial \left( {\left( {{\xi _i} - {u_i}} \right)f} \right)}}{{\partial {\xi _i}}} + R{T_{ES,ij}}\frac{{{\partial ^2}f}}{{\partial {\xi _i}\partial {\xi _j}}}} \right\}.
\end{equation}

\section{Deterministic discrete velocity method for ellipsoidal Fokker-Planck equation}\label{sec:3}
In the DVM framework, the physical space $x_{i}$, the velocity space $\xi_{i}$ and the time $t$ are discrete. The ES-FP equation is solved in an operator splitting way. The free transport operator (LHS of Eq.~\ref{eq:ES-FP}) is solved in the physical space first, in order to get the distribution $f^{*}$ at intermediate step in each discrete element in physical space. Given $f^{*}$, the collision operator (RHS of Eq.~\ref{eq:ES-FP}) can be solved in the discrete velocity space, then the distribution can be evolved to the next time step.

\subsection{Free transport operator}\label{sec:FreeTransprot}
For a numerical scheme in flow predictions, the conservation property in discrete cell (in physical space) is very important when discontinuity, such as the shock wave, exists in the flow field. So, the transport operator of ES-FP equation is solved in a finite volume framework where the extra numerical viscosity needed by capturing the discontinuity is provided by the slope limiters. In this paper, a Euler method is used for temporal discretization. Second order upwind reconstruction in physical space is used for the flux term. The FVM-type numerical scheme for transport operator can be written as
\begin{equation}\label{eq:freetransport1}
\frac{{{f^*} - {f^n}}}{{\Delta t}} + \frac{1}{\Omega }\sum\limits_{a = 1}^A {\left( {{\xi _i}{f_a}} \right){S_{a,i}} = 0},
\end{equation}
where
\begin{equation}\label{eq:freetransport2}
{f_a} = {f^n} + \frac{{\partial {f^n}}}{{\partial {x_j}}}\left( {{x_{a,j}} - {x_{c,j}}} \right).
\end{equation}
In Eq.~\ref{eq:freetransport1}, $S_{a,i}$ is cell interface whose direction is from inside to outside. Its subscript ``$a$" is an index of discrete cell interface, and the total number of discrete interfaces in a cell is denoted by ``$A$". $\Omega$ is the volume of cell. $\Delta t$ is the time interval. The superscript ``$n$" denotes the n-th iteration step, and ``$*$" denotes the intermediate time step between the $n$th and $\left(n+1\right)$th steps in the operator splitting treatment. In Eq.~\ref{eq:freetransport2}, the subscript ``$c$" denotes the ``cell center". In the present method, the calculation of slope $\partial f / \partial x_{j}$ is to the second order, and van Leer slope limiter is used.

\subsection{Collision operator}\label{sec:collision}
The collision operator is solved in a finite difference framework, since it is computational efficient. Theoretically, the evolution equation of collision operator can be directly written using the information at intermediate time step as follows
\begin{equation}
\frac{{{f^{{n + 1}}} - {f^*}}}{{\Delta t}} = \frac{1}{{\tau _{ES}^*}}\left\{ {\frac{{\partial \left( {\left( {{\xi _i} - u_i^*} \right){f^*}} \right)}}{{\partial {\xi _i}}} + RT_{ES,ij}^*\frac{{{\partial ^2}{f^*}}}{{\partial {\xi _i}\partial {\xi _j}}}} \right\},
\end{equation}
where the first and second order slopes in velocity space can be numerically approximated using central difference.

The numerical approximation to the slopes has truncation error related to $\Delta \xi$. If a second order central difference is used, the truncation error is $O\left(\Delta {\xi}^2\right)$. The numerical quadratures in velocity space also generate numerical errors. If the above numerical scheme for ES-FP collision operator is directly used without treatment of these numerical errors, aggregate effect will produce undesired variations of mass, momentum and energy which should be zero since the collision operator fulfills the conservation property. As a result, it often leads to a non-convergent and non-conservative numerical scheme. This problem is addressed in several works in the topic of FPL equation~\cite{Degond1994An,Buet2002Numerical}. For RFP equation which has a similar mathematical form as the FP and ES-FP equations used in gas-kinetic theory, Ref~\cite{Taitano2015A} constructed a conservative finite volume scheme in velocity space.

In this paper, the non-convergence problem for efficient finite difference framework is handled in a similar way as Ref~\cite{Taitano2015A} as follows. Because in a finite difference framework in velocity space, the distribution function only lives at discrete nodes, the corresponding discrete collision operator should be slightly different from the continuous one due to the inevitable numerical errors. First, the collision operator is rewritten as follows by decomposing the advection term into a distribution function and a first order slope,
\begin{equation}
\frac{{\partial f}}{{\partial t}} = \frac{1}{{{\tau _{ES}}}}\left\{ {3f + \left( {{\xi _i} - {u_i}} \right)\frac{{\partial f}}{{\partial {\xi _i}}} + R{T_{ES,ij}}\frac{{{\partial ^2}f}}{{\partial {\xi _i}\partial {\xi _j}}}} \right\}.
\end{equation}
Then each term in the brace is multiplied by a coefficient $\varepsilon$ which is designed to eliminate the influence of numerical errors on conservation property, and the collision operator turns into
\begin{equation}\label{eq:discreteFP}
\frac{{\partial f}}{{\partial t}} = \frac{1}{{{\tau _{ES}}}}\left\{ {{3\varepsilon _F}f + {\varepsilon _{A,i}}\left( {{\xi _i} - {u_i}} \right)\frac{{\partial f}}{{\partial {\xi _i}}} + {\varepsilon _D}R{T_{ES,ij}}\frac{{{\partial ^2}f}}{{\partial {\xi _i}\partial {\xi _j}}}} \right\}.
\end{equation}
The values of $\varepsilon_{F}$, $\varepsilon_{A,i}$ and $\varepsilon_{D}$ are very close to unity, and their departures from unity (denoted by $\left| {\varepsilon  - 1} \right|$) are directly related to the numerical error. Being the similar with Ref.~\cite{Taitano2015A}, the treatment of $\varepsilon _{A,i}$ is as follows,
\begin{equation}
\left\{ {\begin{array}{*{20}{c}}
{{\varepsilon _{A,i}} \ne {\rm{1}},{\rm{     }}{\xi _i} < 0},\\
{{\varepsilon _{A,i}}{\rm{ = 1}},{\rm{     }}{\xi _i} \ge 0}.
\end{array}} \right.
\end{equation}
That means that $\varepsilon _{A,i}$ only exerts on half of the velocity space. For continuous velocity space, these $\varepsilon$ become unity, since it is the basic property of ES-FP equation that mass, momentum and energy conservations are fulfilled. For discrete velocity space, these coefficients can be obtained using the conservation of mass, momentum, and energy, and solving the following algebraic equations analytically,
\begin{equation}\label{eq:calculateEP}
\begin{aligned}
&{\varepsilon _F}\sum 3f  + {\varepsilon _{A,1}}\sum\limits_{{\xi _1} < 0} {{A_1}}  + {\varepsilon _{A,2}}\sum\limits_{{\xi _2} < 0} {{A_2}}  + {\varepsilon _{A,3}}\sum\limits_{{\xi _3} < 0} {{A_3}}  + {\varepsilon _D}\sum {{D}}  = {\rm{ - }}\left( {\sum\limits_{{\xi _1} \ge 0} {{A_1}}  + \sum\limits_{{\xi _2} \ge 0} {{A_2}}  + \sum\limits_{{\xi _3} \ge 0} {{A_3}} } \right),\\
&{\varepsilon _F}\sum {{3\xi _1}f}  + {\varepsilon _{A,1}}\sum\limits_{{\xi _1} < 0} {{\xi _1}{A_1}}  + {\varepsilon _{A,2}}\sum\limits_{{\xi _2} < 0} {{\xi _1}{A_2}}  + {\varepsilon _{A,3}}\sum\limits_{{\xi _3} < 0} {{\xi _1}{A_3}}  + {\varepsilon _D}\sum {{\xi _1}{D}}  = {\rm{ - }}\left( {\sum\limits_{{\xi _1} \ge 0} {{\xi _1}{A_1}}  + \sum\limits_{{\xi _2} \ge 0} {{\xi _1}{A_2}}  + \sum\limits_{{\xi _3} \ge 0} {{\xi _1}{A_3}} } \right),\\
&{\varepsilon _F}\sum {{3\xi _2}f}  + {\varepsilon _{A,1}}\sum\limits_{{\xi _1} < 0} {{\xi _2}{A_1}}  + {\varepsilon _{A,2}}\sum\limits_{{\xi _2} < 0} {{\xi _2}{A_2}}  + {\varepsilon _{A,3}}\sum\limits_{{\xi _3} < 0} {{\xi _2}{A_3}}  + {\varepsilon _D}\sum {{\xi _2}{D}}  = {\rm{ - }}\left( {\sum\limits_{{\xi _1} \ge 0} {{\xi _2}{A_1}}  + \sum\limits_{{\xi _2} \ge 0} {{\xi _2}{A_2}}  + \sum\limits_{{\xi _3} \ge 0} {{\xi _2}{A_3}} } \right),\\
&{\varepsilon _F}\sum {{3\xi _3}f}  + {\varepsilon _{A,1}}\sum\limits_{{\xi _1} < 0} {{\xi _3}{A_1}}  + {\varepsilon _{A,2}}\sum\limits_{{\xi _2} < 0} {{\xi _3}{A_2}}  + {\varepsilon _{A,3}}\sum\limits_{{\xi _3} < 0} {{\xi _3}{A_3}}  + {\varepsilon _D}\sum {{\xi _3}{D}}  = {\rm{ - }}\left( {\sum\limits_{{\xi _1} \ge 0} {{\xi _3}{A_1}}  + \sum\limits_{{\xi _2} \ge 0} {{\xi _3}{A_2}}  + \sum\limits_{{\xi _3} \ge 0} {{\xi _3}{A_3}} } \right),\\
&{\varepsilon _F}\sum {{3\xi ^2}f}  + {\varepsilon _{A,1}}\sum\limits_{{\xi _1} < 0} {{\xi ^2}{A_1}}  + {\varepsilon _{A,2}}\sum\limits_{{\xi _2} < 0} {{\xi ^2}{A_2}}  + {\varepsilon _{A,3}}\sum\limits_{{\xi _3} < {\rm{0}}} {{\xi ^2}{A_3}}  + {\varepsilon _D}\sum {{\xi ^2}{D}}  = {\rm{ - }}\left( {\sum\limits_{{\xi _1} \ge 0} {{\xi ^2}{A_1}}  + \sum\limits_{{\xi _2} \ge 0} {{\xi ^2}{A_2}}  + \sum\limits_{{\xi _3} \ge {\rm{0}}} {{\xi ^2}{A_3}} } \right),
\end{aligned}
\end{equation}
where $A_{i}$ ($i=1,2,3$) and $D$ are defined as
\begin{equation}\label{eq:AD}
\begin{aligned}
&{A_i} = \left( {{\xi _i} - {u_i}} \right)\frac{{\partial f}}{{\partial {\xi _i}}},\\
&D = R{T_{ES,ij}}\frac{{{\partial ^2}f}}{{\partial {\xi _i}\partial {\xi _j}}}.
\end{aligned}
\end{equation}
In Eq.\ref{eq:calculateEP}, the numerical integrals in whole velocity space and half velocity space are denoted by $\sum {\left(  \cdot  \right)}$, $\sum\limits_{{\xi _i} < 0} {\left(  \cdot  \right)}$, and $\sum\limits_{{\xi _i} \ge 0} {\left(  \cdot  \right)}$ respectively. Take index $i=1$ for example, these integrals are in the following form
\begin{equation}\label{eq:sum}
\begin{aligned}
&\sum {\left(  \cdot  \right)}  = \sum\limits_{{\xi _1} \in \left( {{\xi _{1,\min }},{\xi _{1,\max }}} \right)} {\sum\limits_{{\xi _2} \in \left( {{\xi _{2,\min }},{\xi _{2,\max }}} \right)} {\sum\limits_{{\xi _3} \in \left( {{\xi _{3,\min }},{\xi _{3,\max }}} \right)} {\left(  \cdot  \right)\Delta {\xi _1}\Delta {\xi _2}\Delta {\xi _3}} } }, \\
&\sum\limits_{{\xi _1} < 0} {\left(  \cdot  \right)}  = \sum\limits_{{\xi _1} \in \left( {{\xi _{1,\min }},{\rm{0}}} \right)} {\sum\limits_{{\xi _2} \in \left( {{\xi _{2,\min }},{\xi _{2,\max }}} \right)} {\sum\limits_{{\xi _3} \in \left( {{\xi _{3,\min }},{\xi _{3,\max }}} \right)} {\left(  \cdot  \right)\Delta {\xi _1}\Delta {\xi _2}\Delta {\xi _3}} } },\\
&\sum\limits_{{\xi _1} \ge 0} {\left(  \cdot  \right)}  = \sum\limits_{{\xi _1} \in \left[ {\left. {0,{\xi _{1,\max }}} \right)} \right.} {\sum\limits_{{\xi _2} \in \left( {{\xi _{2,\min }},{\xi _{2,\max }}} \right)} {\sum\limits_{{\xi _3} \in \left( {{\xi _{3,\min }},{\xi _{3,\max }}} \right)} {\left(  \cdot  \right)\Delta {\xi _1}\Delta {\xi _2}\Delta {\xi _3}} } }.
\end{aligned}
\end{equation}
where the subscript ``$min$" and ``$max$" denote the boundaries in each direction of truncated velocity space.

Finally, the evolution of distribution function according to discrete ES-FP collision operator can be written as
\begin{equation}\label{eq:FPevolution}
\frac{{{f^{{n + 1}}} - {f^*}}}{{\Delta t}} = \frac{1}{{\tau _{ES}^*}}\left\{ { {3\varepsilon _F^*} {f^*} +  {\varepsilon _{A,i}^*} \left( {{\xi _i} - u_i^*} \right)\frac{{\partial {f^*}}}{{\partial {\xi _i}}} +  {\varepsilon _F^*} RT_{ES,ij}^*\frac{{{\partial ^2}{f^*}}}{{\partial {\xi _i}\partial {\xi _j}}}} \right\}.
\end{equation}

The calculation process of collision operator can be summed up as follows. First, using the information $f^{*}$ at the intermediate step, the first and second order slopes (in velocity space) in Eq.~\ref{eq:FPevolution} can be calculated and stored. Then the contribution of each term in the brace of Eq.~\ref{eq:FPevolution} to mass, momentum, and energy can be calculated using numerical integration in Eq.~\ref{eq:sum}. In the process of numerical integration, $u_i$ and $T_{ij}$ can also be obtained as
\begin{equation}
\begin{aligned}
&T_{ij}^* = \frac{{\sum {{c_i}{c_j}{f^*}} }}{R{\sum {{f^*}} }},\\
&u_i^* = \frac{{\sum {{\xi _i}{f^*}} }}{{\sum {{f^*}} }}.
\end{aligned}
\end{equation}
Then, using the obtained numerical integrals, the coefficient $\varepsilon^{*}$ can be calculated using Eq.~\ref{eq:calculateEP}. Up to this point, every term in Eq.~\ref{eq:FPevolution} is obtained, and the distribution function can be updated to the $\left(n+1\right)$th time step.

\subsection{Reduced ES-FP equation}
Real monatomic gas flows have a three dimensional physical space and a three dimensional velocity space ($3D$-$3V$ case). For $nD$-$3V$ case where $n<3$, the ES-FP equation can be reduced to $nD$-quasi $nV$ case whose computational cost is greatly reduced. The following context takes the reducing process from $1D$-$3V$ case to $1D$-quasi $1V$ case for example (which is also used in the case of shock wave structure calculation in this paper). For $1D$ case in $x_1$ direction, there is
\begin{equation}
\begin{aligned}
&\frac{{\partial f}}{{\partial {x_2}}} = 0,{\rm{   }}\frac{{\partial f}}{{\partial {x_3}}} = 0,\\
&u_{2} = 0,{\rm{   }}u_{3} = 0,\\
&{T_{12}} = {T_{21}} = 0,{\rm{   }}{T_{23}} = {T_{32}} = 0,{\rm{   }}{T_{13}} = {T_{31}} = 0,\\
&{\rm{   }}{T_{22}} = {T_{{\rm{33}}}}{\rm{ = }}\left( {3T - {T_{11}}} \right)/2.\\
\end{aligned}
\end{equation}
The slopes in $x_2$ and $x_3$ directions are zero. The off-diagonal elements in temperature tensor $T_{ij}$ are zero because tangential stress is zero (according to Eq.~\ref{eq:constrain}).

ES-FP equation can be first reduced to
\begin{equation}
\frac{{\partial f}}{{\partial t}} + {\xi _1}\frac{{\partial f}}{{\partial {x_1}}} = \frac{1}{{{\tau _{ES}}}}\left\{ {3f + \left( {{\xi _1} - {u_1}} \right)\frac{{\partial f}}{{\partial {\xi _1}}} + {{\xi _2}}\frac{{\partial f}}{{\partial {\xi _2}}}+ {{\xi _3}}\frac{{\partial f}}{{\partial {\xi _3}}} + R{T_{ES,11}}\frac{{{\partial ^2}f}}{{\partial \xi _1^2}} + R{T_{ES,22}}\frac{{{\partial ^2}f}}{{\partial \xi _2^2}} + R{T_{ES,33}}\frac{{{\partial ^2}f}}{{\partial \xi _3^2}}} \right\}.
\end{equation}
Define a mass distribution and an energy distribution in $\xi_1$ axis as follows
\begin{equation}
\begin{aligned}
&F = \int_{ - \infty }^{ + \infty } {\int_{ - \infty }^{ + \infty } {mfd{\xi _2}d{\xi _3}} },\\
&G = \int_{ - \infty }^{ + \infty } {\int_{ - \infty }^{ + \infty } {m\left( {\xi _2^2 + \xi _3^2} \right)fd{\xi _2}d{\xi _3}} }.
\end{aligned}
\end{equation}
Then multiply ES-FP equation by unity and $\xi_{2}^2+\xi_{3}^2$, and integrate it in both $\xi_{2}$ and $\xi_{3}$ directions. After calculating the integrals, the reduced ES-FP equation becomes a system of two equations about ``$F$" and ``$G$" as follows
\begin{equation}\label{eq:FG}
\begin{aligned}
&\frac{{\partial F}}{{\partial t}} + {\xi _1}\frac{{\partial F}}{{\partial {x_1}}} = \frac{1}{{{\tau _{ES}}}}\left\{ {F + \left( {{\xi _1} - {u_1}} \right)\frac{{\partial F}}{{\partial {\xi _1}}} + R{T_{ES,11}}\frac{{{\partial ^2}F}}{{\partial \xi _1^2}}} \right\},\\
&\frac{{\partial G}}{{\partial t}} + {\xi _1}\frac{{\partial G}}{{\partial {x_1}}} = \frac{1}{{{\tau _{ES}}}}\left\{ {G + \left( {{\xi _1} - {u_1}} \right)\frac{{\partial G}}{{\partial {\xi _1}}} - 2G + R{T_{ES,11}}\frac{{{\partial ^2}G}}{{\partial \xi _1^2}} + 2R\left( {{T_{ES,22}}{\rm{ + }}{T_{ES,{\rm{33}}}}} \right)F} \right\}.
\end{aligned}
\end{equation}
Then the solving of $f$ in three dimensional velocity space is turned into the solving of $F$ and $G$ in one dimensional velocity space. Here the evolutions of $F$ and $G$ are coupled through a relaxation process from $2G$ to $2R\left( {{T_{ES,22}}+{T_{ES,{\rm{33}}}}} \right)F$ in the second equation ($G$ equation) in Eq.~\ref{eq:FG}. If the integrated equation is Eq.~\ref{eq:discreteFP} (the discrete form), $\varepsilon$ will appear in the corresponding terms in both $F$ and $G$ equations in Eq.~\ref{eq:FG}. Practically, these $\varepsilon$ can only appear in $F$ equation, since it can be seen from later equation (Eq.~\ref{eq:1depsilon}) that both mass and momentum conservations are only involved by $F$, and $F$ also appears in the expression of energy conservation. So the reduced ES-FP equations in discrete velocity space can be written as
\begin{equation}\label{eq:FG2}
\begin{aligned}
&\frac{{\partial F}}{{\partial t}} + {\xi _1}\frac{{\partial F}}{{\partial {x_1}}} = \frac{1}{{{\tau _{ES}}}}\left\{ {{\varepsilon _F}F + {\varepsilon _{A,1}}\left( {{\xi _1} - {u_1}} \right)\frac{{\partial F}}{{\partial {\xi _1}}} + {\varepsilon _D}R{T_{ES,11}}\frac{{{\partial ^2}F}}{{\partial \xi _1^2}}} \right\},\\
&\frac{{\partial G}}{{\partial t}} + {\xi _1}\frac{{\partial G}}{{\partial {x_1}}} = \frac{1}{{{\tau _{ES}}}}\left\{ {G + \left( {{\xi _1} - {u_1}} \right)\frac{{\partial G}}{{\partial {\xi _1}}} - 2G + R{T_{ES,11}}\frac{{{\partial ^2}G}}{{\partial \xi _1^2}} + 2R\left( {{T_{ES,22}}{\rm{ + }}{T_{ES,{\rm{33}}}}} \right)F} \right\},
\end{aligned}
\end{equation}
here, in $1D$ case, the ``$A$" and ``$D$" in Eq.~\ref{eq:AD} are reduced to
\begin{equation}
\begin{aligned}
&{A_1} = \left( {{\xi _1} - {u_1}} \right)\frac{{\partial F}}{{\partial {\xi _1}}},\\
&D = R{T_{ES,11}}\frac{{{\partial ^2}F}}{{\partial {\xi _1}\partial {\xi _1}}}.
\end{aligned}
\end{equation}

Using conservations of mass, momentum and energy, the $\varepsilon$ for conservation purpose can be obtained by solving the following linear equations,
\begin{equation}\label{eq:1depsilon}
\begin{aligned}
&{\varepsilon _F}\sum F  + {\varepsilon _{A,1}}\sum\limits_{{\xi _1} < 0} {{A_1}}  +  + {\varepsilon _D}R{T_{ES,11}}\sum {{D_{11}}}  = {\rm{ - }}\left( {\sum\limits_{{\xi _1} < 0} {{A_1}} } \right),\\
&{\varepsilon _F}\sum {{\xi _1}F}  + {\varepsilon _{A,1}}\sum\limits_{{\xi _1} < 0} {{\xi _1}{A_1}}  + {\varepsilon _D}R{T_{ES,11}}\sum {{\xi _1}{D_{11}}}  = {\rm{ - }}\left( {\sum\limits_{{\xi _1} < 0} {{\xi _1}{A_1}} } \right),\\
&{\varepsilon _F}\sum {{\xi_{1}^2}F}  + {\varepsilon _{A,1}}\sum\limits_{{\xi _1} < 0} {{\xi_{1}^2}{A_1}}  + {\varepsilon _D}R{T_{ES,11}}\sum {{\xi_{1}^2}{D_{11}}}  =  - \left( {\sum\limits_{{\xi _1} < 0} {{\xi_{1}^2}{A_1} + \sum {RHS_G} } } \right),
\end{aligned}
\end{equation}
here $RHS_G$ is the RHS of the $G$ equation (Eq.~\ref{eq:FG2}). The numerical process for $1V$ case is the same with the $3V$ case in Sec.~\ref{sec:FreeTransprot} and Sec.~\ref{sec:collision}, expect that the operation of $f$ is now on $F$ and $G$.

\subsection{the numerical error in discrete Fokker-Planck collision operator}
The numerical error in discrete velocity space comes from the three items below:
\begin{enumerate}
  \item the truncation in velocity space,
  \item the error in numerical integration,
  \item the truncation error in calculating the slopes using discrete points.
\end{enumerate}

For the 1st item, the domain of truncated velocity space should be as large as possible. But in order to achieve high computational efficiency, it can not be too large. The Maxwellian distribution suggests that the domain should be at least larger than $3\sqrt{RT}$, since beyond $3\sqrt{RT}$, the distribution only contributes $0.3\%$ of mass.

For the 2nd item, high order numerical integration can be used, such as Newton-Cotes integration, to suppress the numerical error in this item. While for the sake of clarity, the rectangular integration is used in this paper.

For the 3rd item, high order central difference can be used, which will be analyzed in Sec.~\ref{sec:maintain}. Much of its influence is on the order of the departure of $\varepsilon$ from unity. A too high order central difference will harm the computational efficiency.

\section{Numerical experiment}\label{sec:4}
\subsection{Maintain the thermal equilibrium Maxwellian distribution ($0D$-$3V$ case)}\label{sec:maintain}
In this case, the initial distribution function is the Maxwellian distribution $g$ in the following form,
\begin{equation}\label{eq:Maxwell}
g = n{\left( {\frac{m}{{2\pi kT}}} \right)^{{3 \mathord{\left/
 {\vphantom {3 2}} \right.
 \kern-\nulldelimiterspace} 2}}}\exp \left( { - \frac{{m{c_i}{c_i}}}{{2kT}}} \right).
\end{equation}
If a discrete numerical method is conservative, the distribution function will maintain the Maxwellian distribution.

\begin{figure}
\centering
\includegraphics[width=0.45\textwidth]{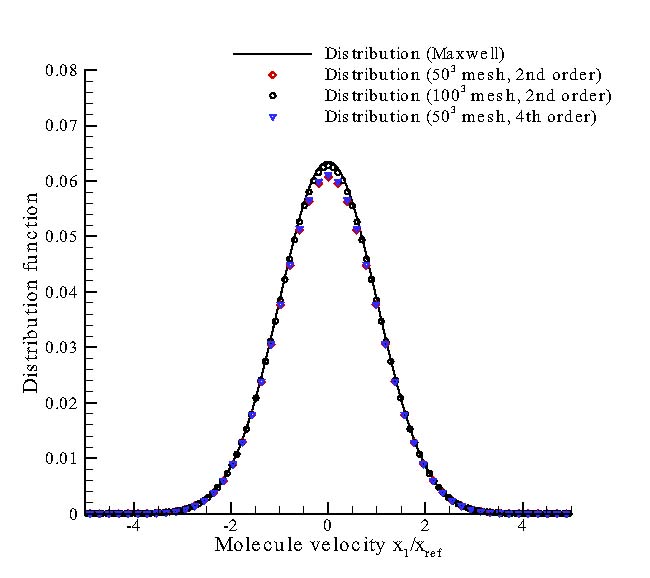}
\caption{\label{Fig:case_stable} Stable Maxwellian distribution computed by FP-DVM on $50^3$ and $100^3$ uniform meshes.}
\end{figure}

Using this case, the numerical stability of the present method is examined. The validity of conservative coefficients $\varepsilon$ and the accuracy of the present scheme are also investigated. For Maxwellian distribution, $n=1$, $u_i=0$, and $T=1$ are chosen. The domain of truncated velocity space in each direction is $[-5,5]$. Both $50*50*50$ and $100*100*100$ meshes in velocity space are tested. Both 2nd and 4rd order central difference are used in approximating the 1st and 2nd order slopes in ES-FP collision operator. The maintained distribution functions that are predicted using $50*50*50$ and $100*100*100$ meshes are shown in radial direction in Fig.~\ref{Fig:case_stable}, respectively. Although the distribution on $50*50*50$ meshes slightly deviates from the analytical Maxwellian distribution near the zero point, it is stable since the discrete ES-FP is conservative. In Table~\ref{table:Maxwell}, there is a comparison of integral error and $\varepsilon$ under dense/coarse meshes and using low/high order central difference for calculating slopes. It can be seen that integral error ($\left|\rho-1\right|$, $\left|u_{1}\right|$, $\left|T-1\right|$) is related to the mesh number, and almost has no relation to the order of central difference. Comparing with the coarse mesh, by using a dense mesh, the precision of macroscopic variables will increase, while the computational cost will also increase. To increase the integration precision, high order integration method such as Newton-Cotes can be used without using a dense mesh. While for clarity, the direct rectangular integration is used in this paper. Using either a dense mesh or a higher order difference, will make the deviation of $\varepsilon$ from unity a smaller value. The largest deviation comes from $\varepsilon_{D}$. From the $log$ data, it can be seen that when 4rd order difference is used, the order of $\left|\varepsilon_{D}-1\right|$ is about 4, while for a 2nd order difference, it is about 2. This data shows that the order of $\left|\varepsilon_{D}-1\right|$ is related to the order of numerical difference, and is almost not affected by the mesh number. In the following test cases, the second order central difference is used in the velocity space for efficiency.

\subsection{Energy relaxation among directions ($0D$-$3V$ case)}
In this case, initially the temperatures in different directions are not the same (anisotropic temperature). Through molecular collisions, these temperatures will gradually achieve equilibrium during several m.c.t.. Here the initial temperatures are set to be $T_{1} = 2.0$, $T_{2} = 1.0$, $T_{3} = 1.0$. The truncated discrete velocity space is $[-7,7]$ in each direction with 70 discrete points (70*70*70 mesh). The iteration time $\Delta t$ is chosen as $0.005\tau$. Here $\tau=\mu/p$ has the same order of magnitude as $\tau_{FP}$ and $\tau_{ES}$, but their values are not the same.

\begin{figure}
\centering
\subfigure[$\tau=0$]{
\includegraphics[width=0.45\textwidth]{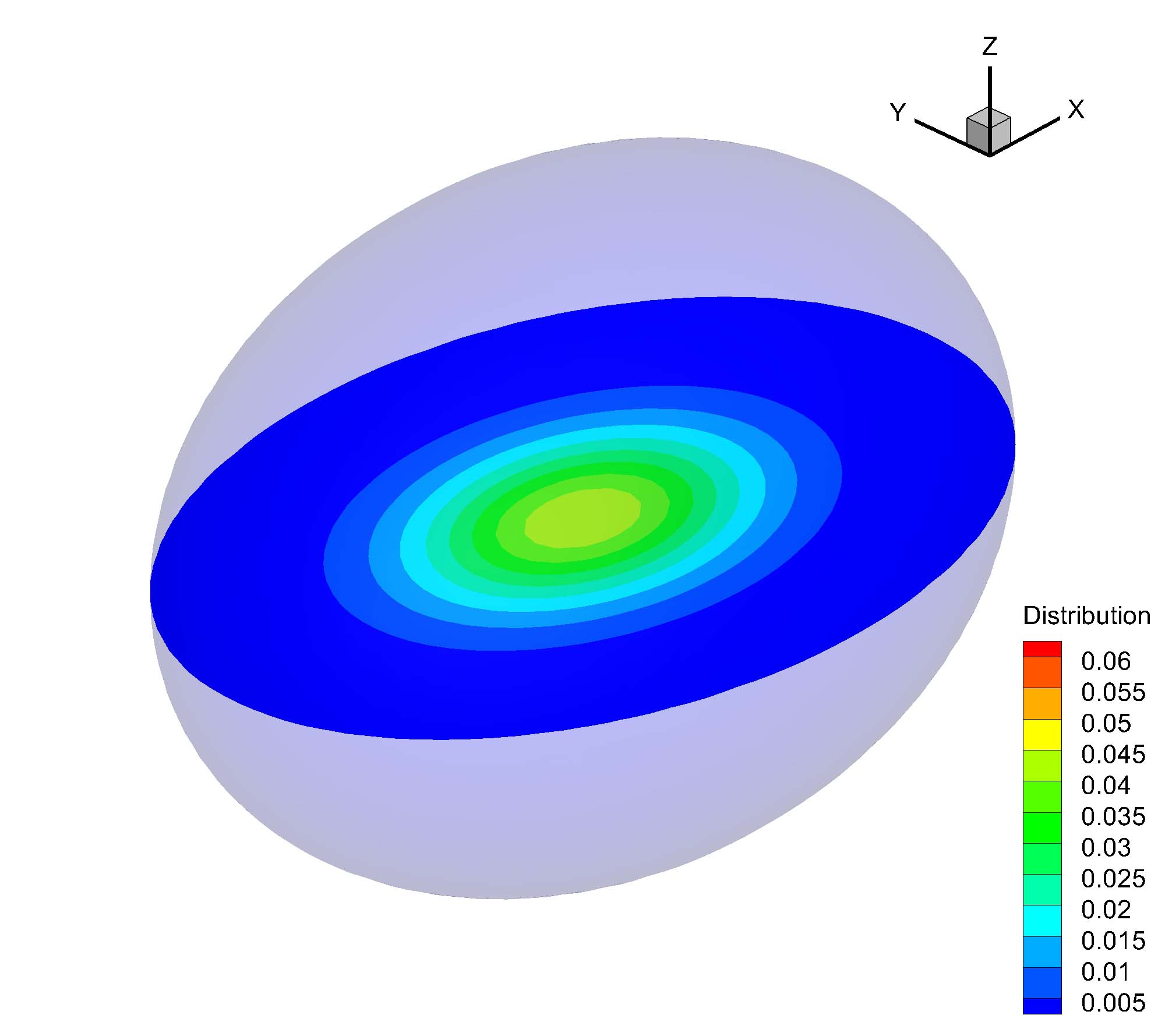}
}\hspace{0.05\textwidth}
\subfigure[$\tau=1$]{
\includegraphics[width=0.45\textwidth]{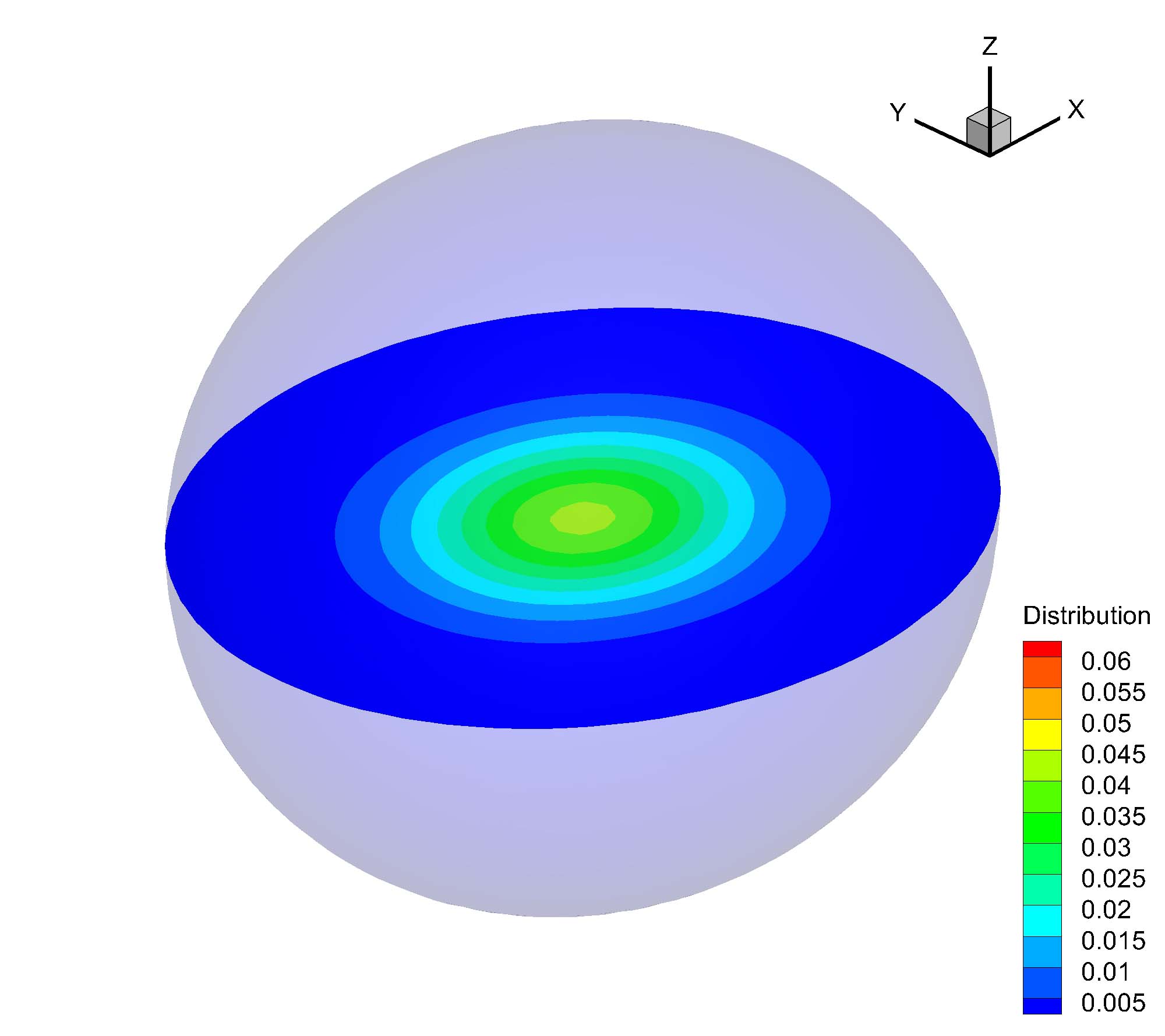}
}
\subfigure[$\tau=2$]{
\includegraphics[width=0.45\textwidth]{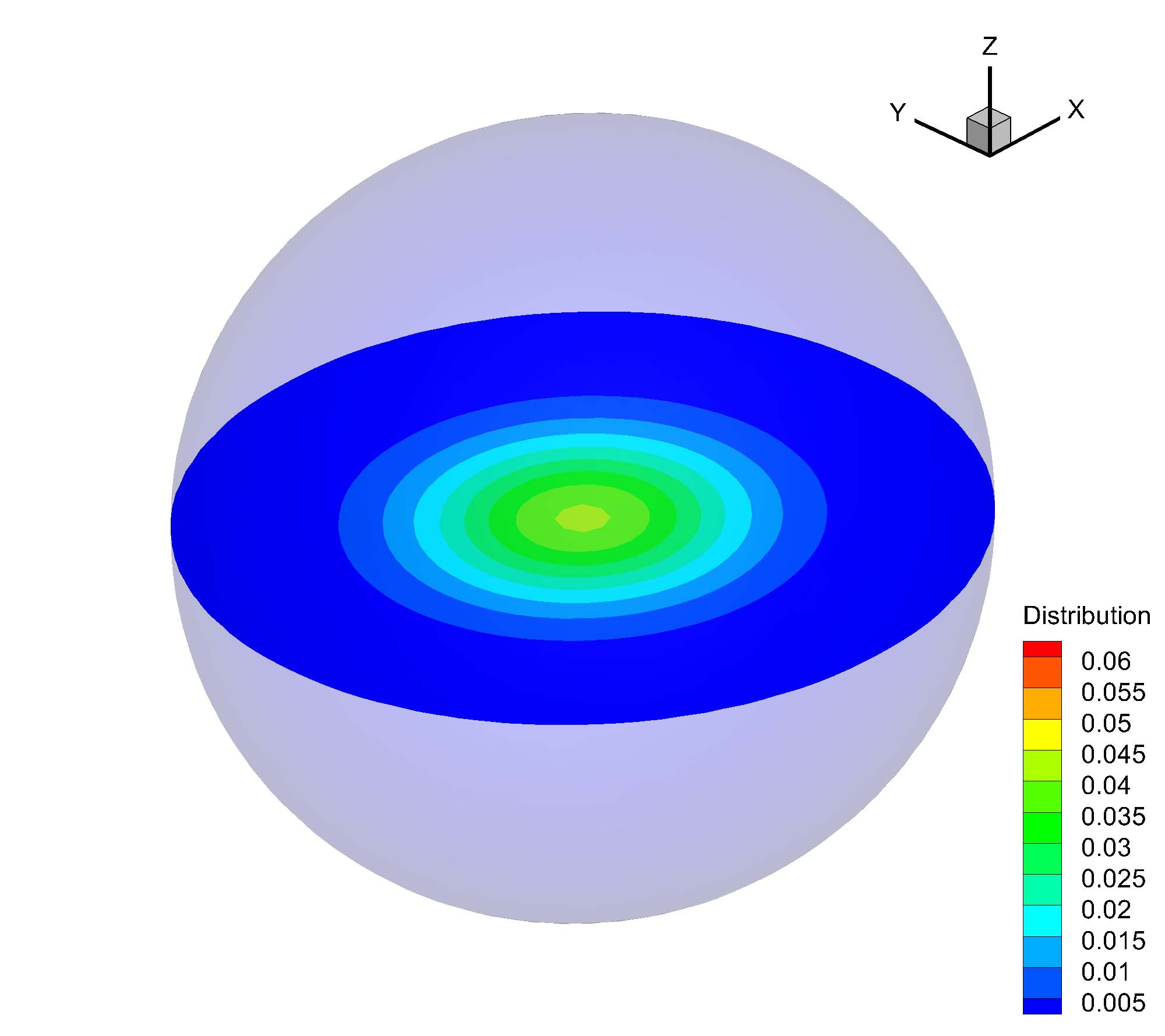}
}
\subfigure[$\tau=10$]{
\includegraphics[width=0.45\textwidth]{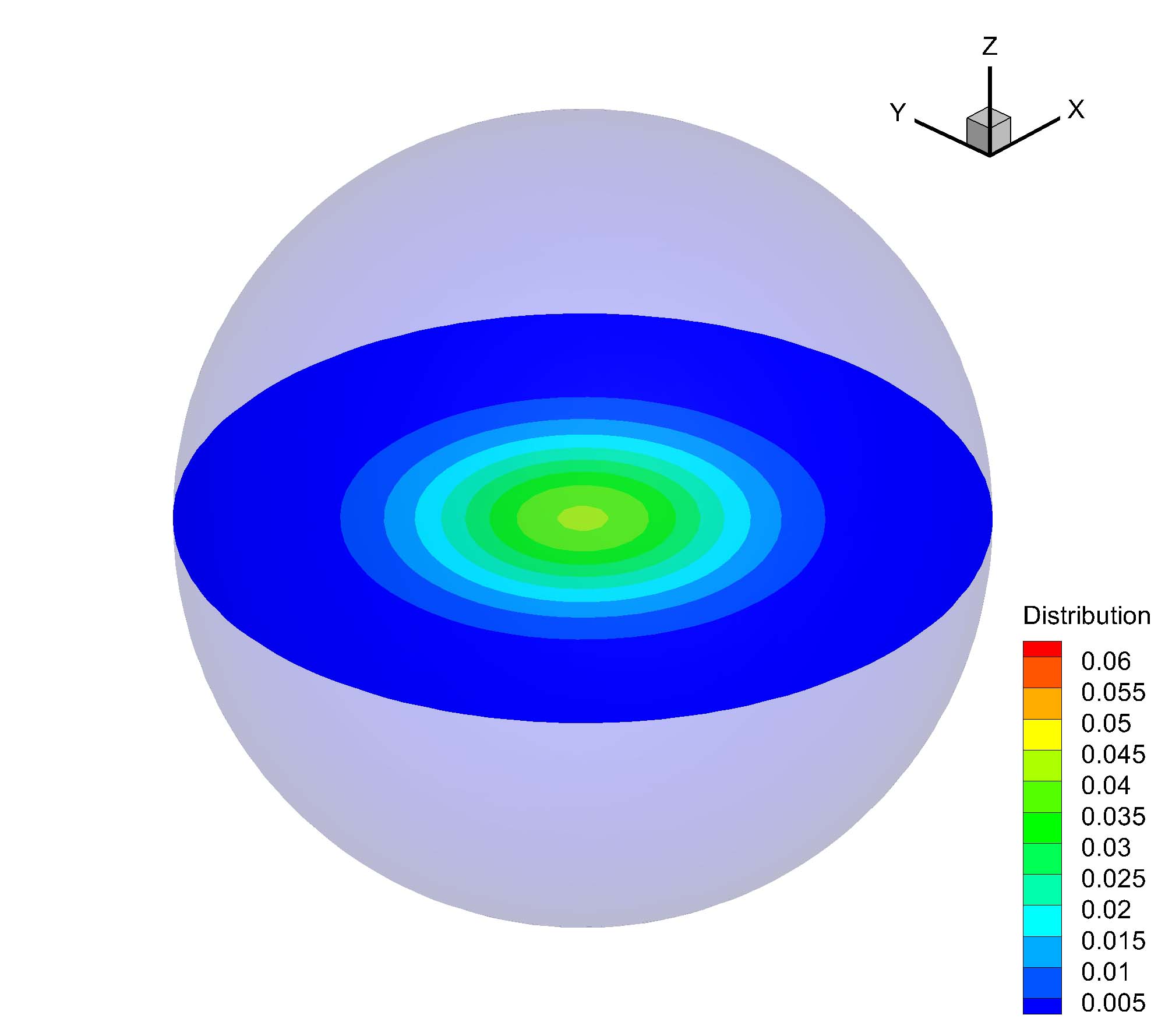}
}
\caption{\label{Fig:anisotropic_distribution} The relaxation process of distribution with initial anisotropic temperatures, only the region with distribution function greater than $5\times10^{-3}$ is plotted, and the contour is on the plane $\xi_{3}=0$.}
\end{figure}

\begin{figure}
\centering
\includegraphics[width=0.45\textwidth]{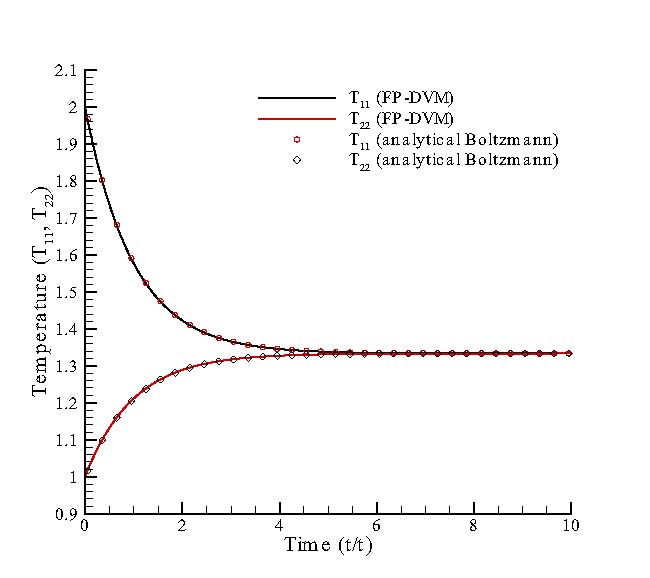}
\caption{\label{Fig:case_unisotropic_relaxation} The relaxation process of anisotropic temperatures with initial values $T_{11} = 2T_{22}=2T_{33}$.}
\end{figure}

For Maxwell molecule, the relaxation of stress and heat flux from the Boltzmann equation is~\cite{Cercignani1990The}
\begin{equation}\label{eq:tauq}
\begin{aligned}
&\frac{{\partial {\tau _{ij}}}}{{\partial t}} =  - \frac{{{\tau _{ij}}}}{\tau },\\
&\frac{{\partial {q_i}}}{{\partial t}} =  - \frac{{\Pr {q_i}}}{\tau }.
\end{aligned}
\end{equation}
For homogenous case, since the density is a constant, the relaxation of anisotropic temperature can be derived from Eq.~\ref{eq:tauq} as follows
\begin{equation}
\frac{{\partial {T_{ij}}}}{{\partial t}} =  - \frac{{{T_{ij}} - T{\delta _{ij}}}}{\tau }.
\end{equation}
So, the analytical solution of temperature and heat flux can be obtained as
\begin{equation}\label{eq:analytical Boltzmann}
\begin{aligned}
&{T_{ij}}\left( t \right) = {e^{ - t/\tau }}\left\{ {{T_{ij}}\left( 0 \right) - T\left( 0 \right){\delta _{ij}}} \right\} + T\left( 0 \right){\delta _{ij}},\\
&q_{i}\left( t \right) = {e^{ - \Pr t/\tau }}q_{i}\left( 0 \right).
\end{aligned}
\end{equation}
The relaxation process of distribution in $3V$ space predicted by the present method is shown in Fig.~\ref{Fig:anisotropic_distribution}, where the iso-surface of distribution gradually transforms from an ellipsoid to a sphere during several $\tau$. In Fig.~\ref{Fig:case_unisotropic_relaxation}, the relaxation process of anisotropic temperatures predicted by the present method matches precisely with the analytical solution (Eq.~\ref{eq:analytical Boltzmann}).

\subsection{Relaxation of bi-model distribution function ($0D$-$3V$ case)}
\begin{figure}
\centering
\subfigure[$\tau=0$]{
\includegraphics[width=0.45\textwidth]{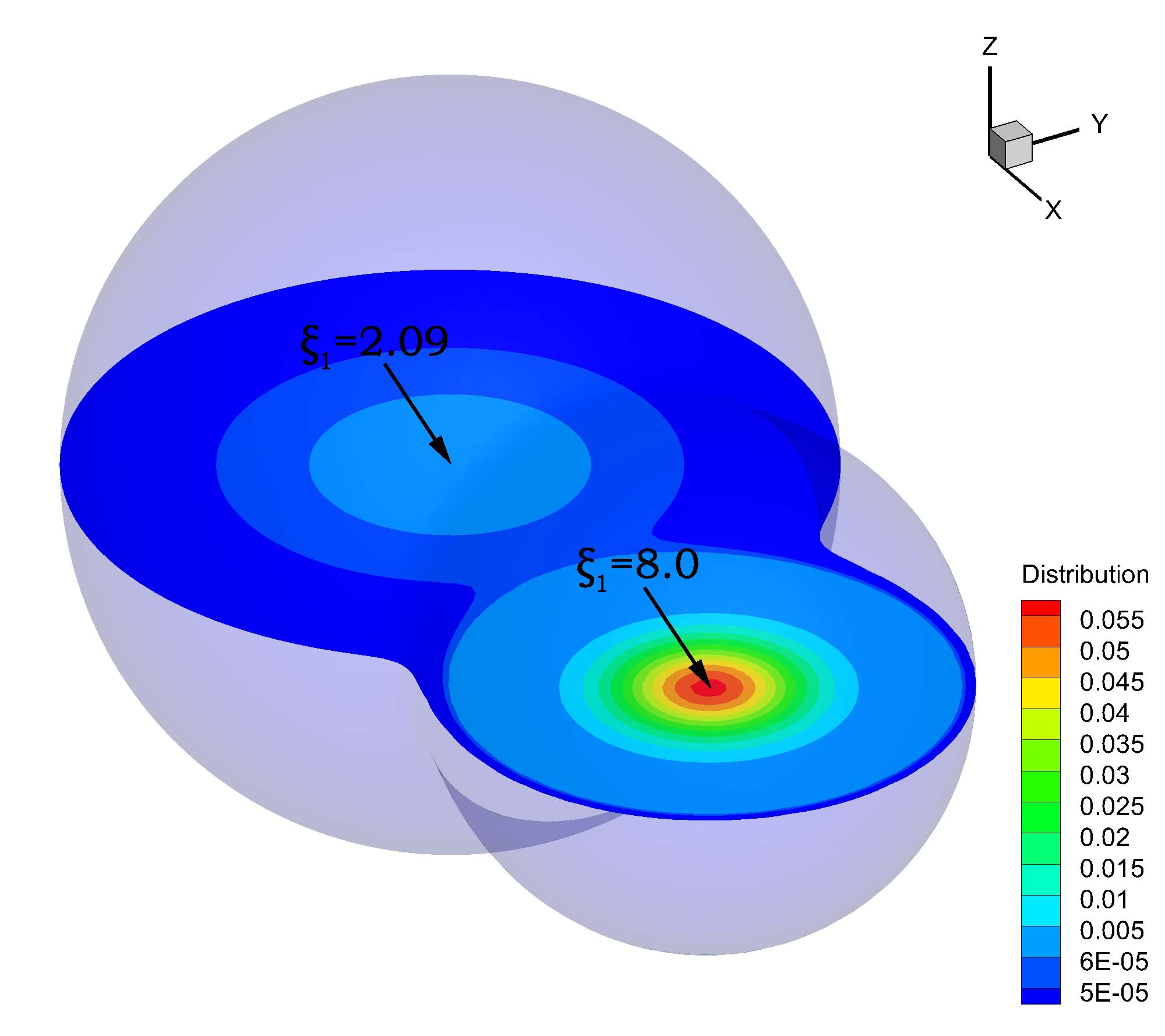}
}\hspace{0.05\textwidth}
\subfigure[$\tau=1$]{
\includegraphics[width=0.45\textwidth]{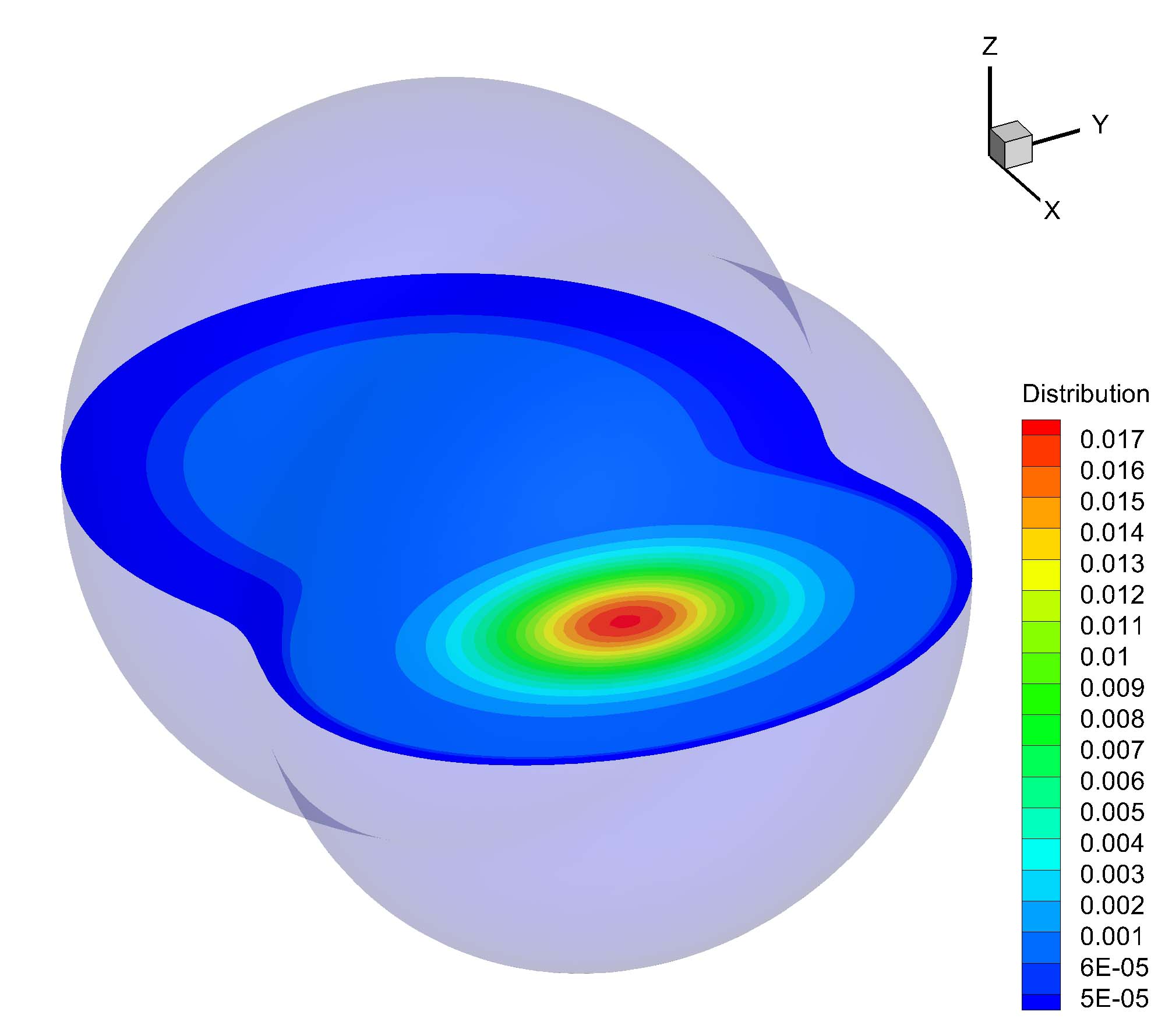}
}
\subfigure[$\tau=2$]{
\includegraphics[width=0.45\textwidth]{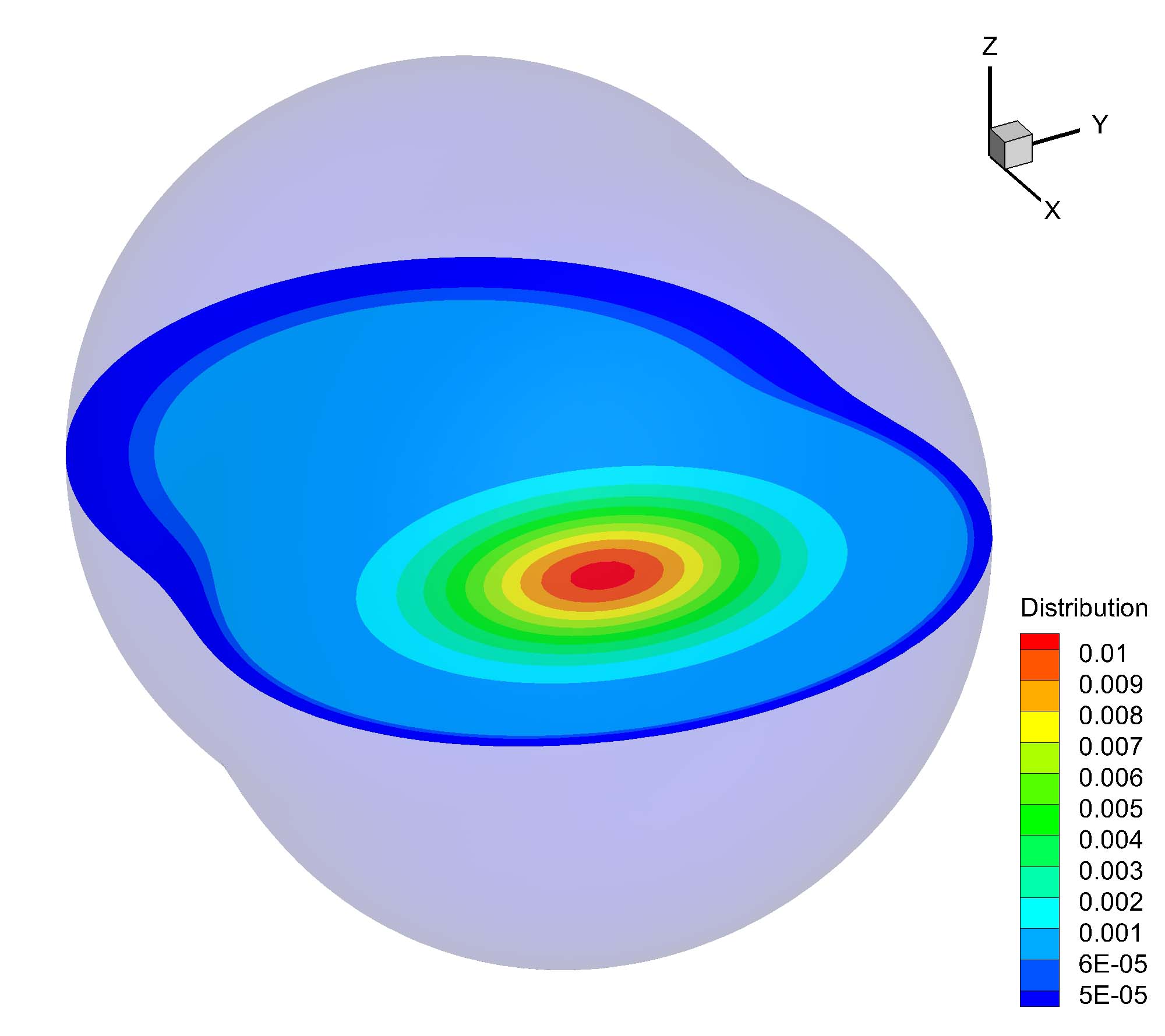}
}
\subfigure[$\tau=10$]{
\includegraphics[width=0.45\textwidth]{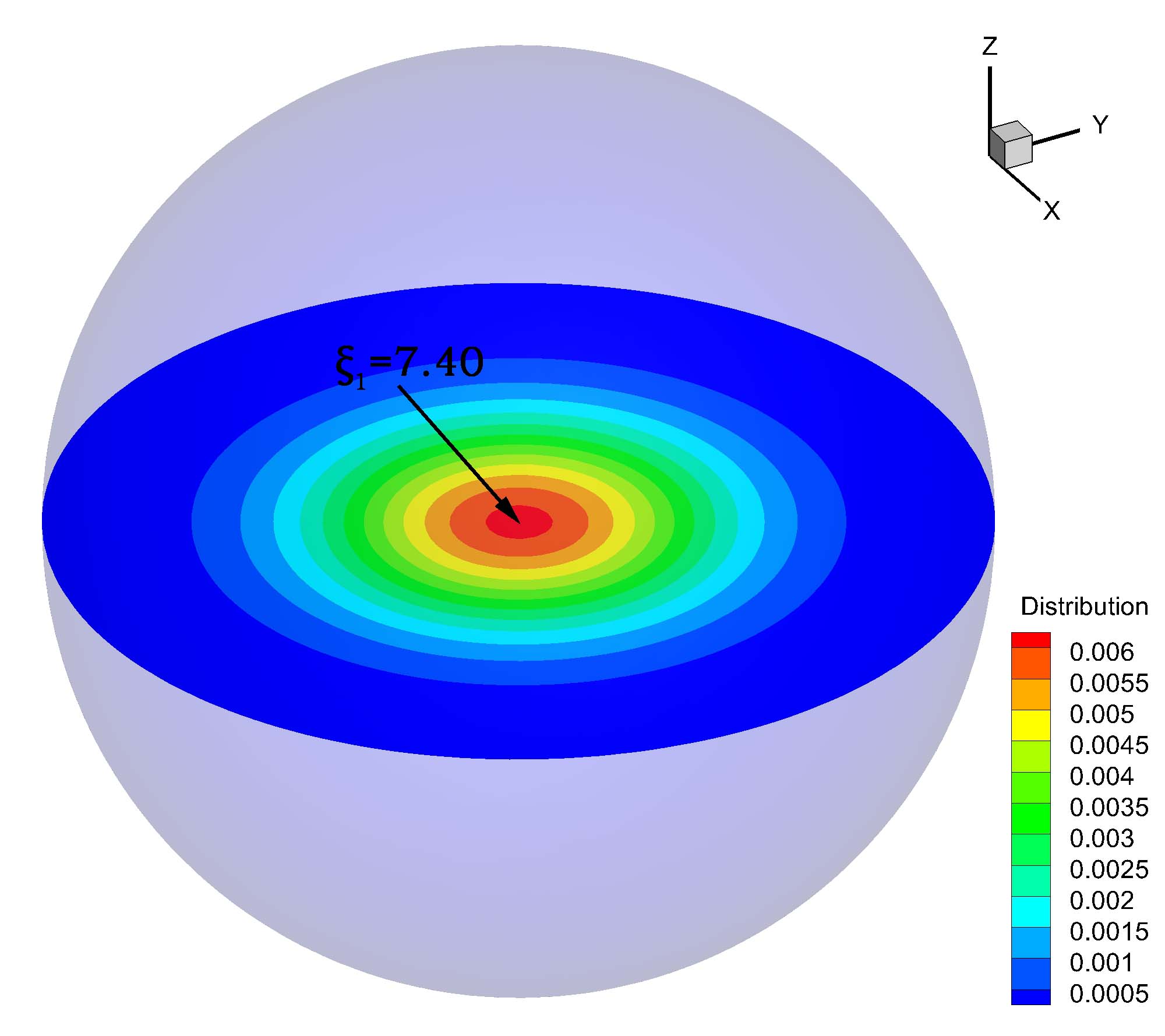}
}
\caption{\label{Fig:bi-model distribution} The relaxation process of bi-model distribution, only the region with distribution function greater than $5\times10^{-5}$ (for subfigure a-c) or $5\times10^{-4}$ (for subfigure d) is plotted, and the contour is on the plane $\xi_{3}=0$.}
\end{figure}

\begin{figure}
\centering
\subfigure[\label{Fig:case_shock_relaxationa} anisotropic temperatures]{
\includegraphics[width=0.45\textwidth]{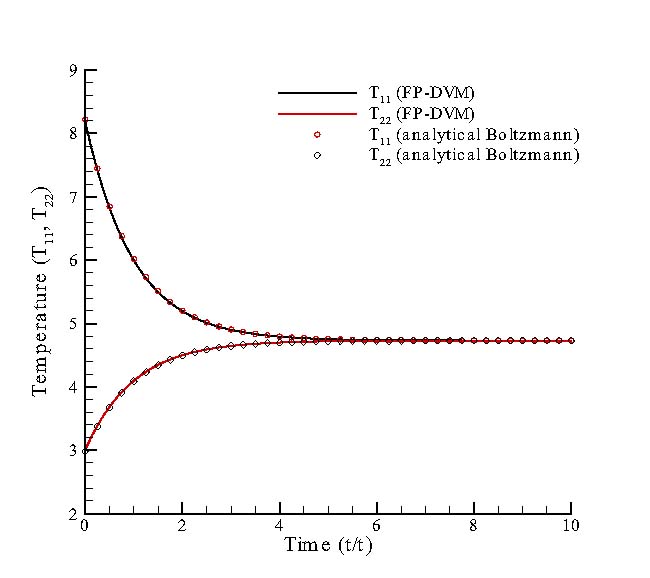}
}\hspace{0.05\textwidth}
\subfigure[\label{Fig:case_shock_relaxationb} heat flux]{
\includegraphics[width=0.45\textwidth]{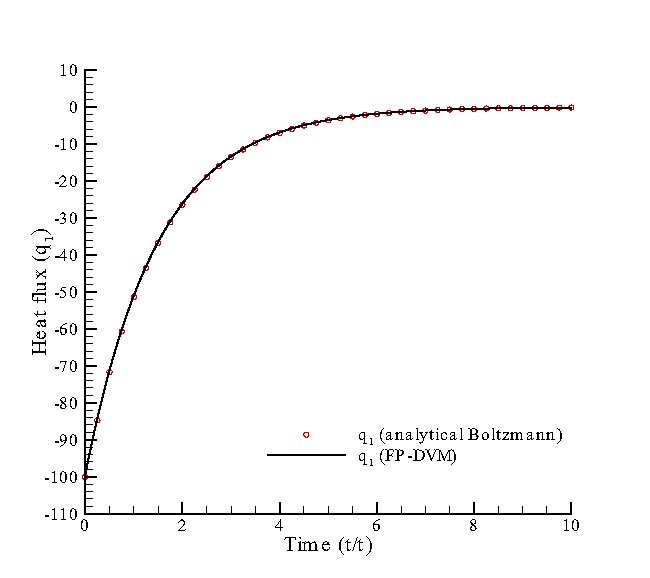}
}
\caption{\label{Fig:case_shock_relaxation} The relaxation process of anisotropic temperatures and heat flux of initial bi-model distribution.}
\end{figure}

In this case, the distribution function is composed of two Maxwellian distributions determined by the physical variables before and after the shock wave respectively. According to the Rankine-Hugoniot relation for a $Mach$ 8.0 shock wave, the physical variables before the shock are $u_{a,1}=8.0$, $u_{a,2}=u_{a,3}=0$, $T_{a}=1.0$, and the physical variables after the shock wave are $u_{b,1}=2.09$, $u_{b,2}=u_{b,3}=0$, $T_{b}=20.87$. The weights of two Maxwellian distributions are chosen as $\rho_a=0.9$ and $\rho_b=0.1$, in order to mimic the distribution function in the front of shock wave, where high non-equilibrium exists. This case investigates the relaxation of this highly non-equilibrium distribution function. The truncated discrete velocity space is $[-26,26]$ in each direction with 260 discrete points (260*260*260 mesh). The iteration time $\Delta t$ is chosen as $0.001\tau$. The time evolution of this initial bi-model distribution is shown in Fig.~\ref{Fig:bi-model distribution}, where two Maxwellian distributions merge into a single one during about $10\tau$. The evolutions of anisotropic temperatures and heat flux predicted by the present method are shown in Fig.\ref{Fig:case_shock_relaxation}, and they match with the analytical solution (Eq.~\ref{eq:analytical Boltzmann}) precisely.

\subsection{Relaxation of discontinuous distribution function ($0D$-$3V$ case)}
\begin{figure}
\centering
\subfigure[$\tau=0$]{
\includegraphics[width=0.45\textwidth]{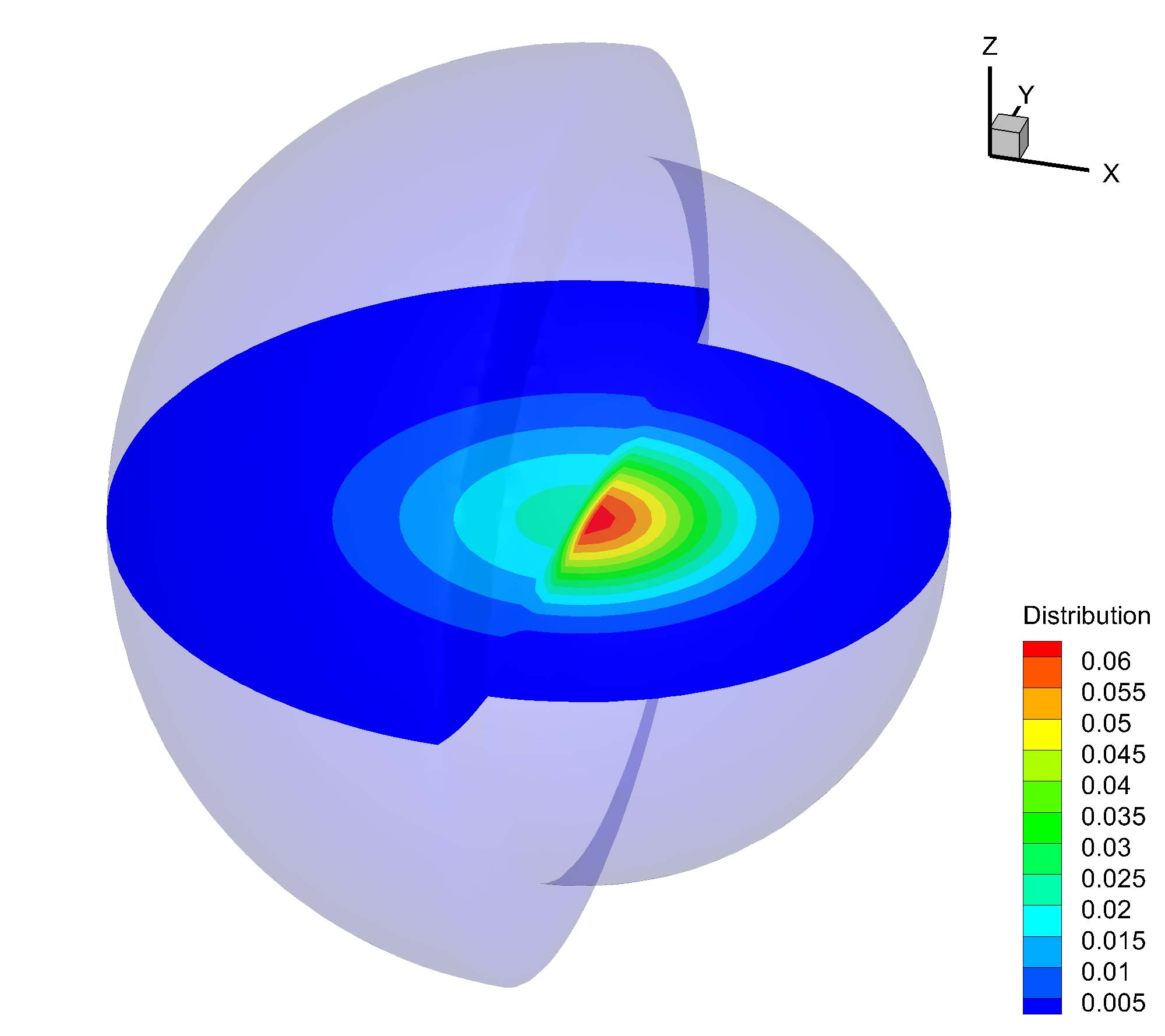}
}\hspace{0.05\textwidth}
\subfigure[$\tau=1$]{
\includegraphics[width=0.45\textwidth]{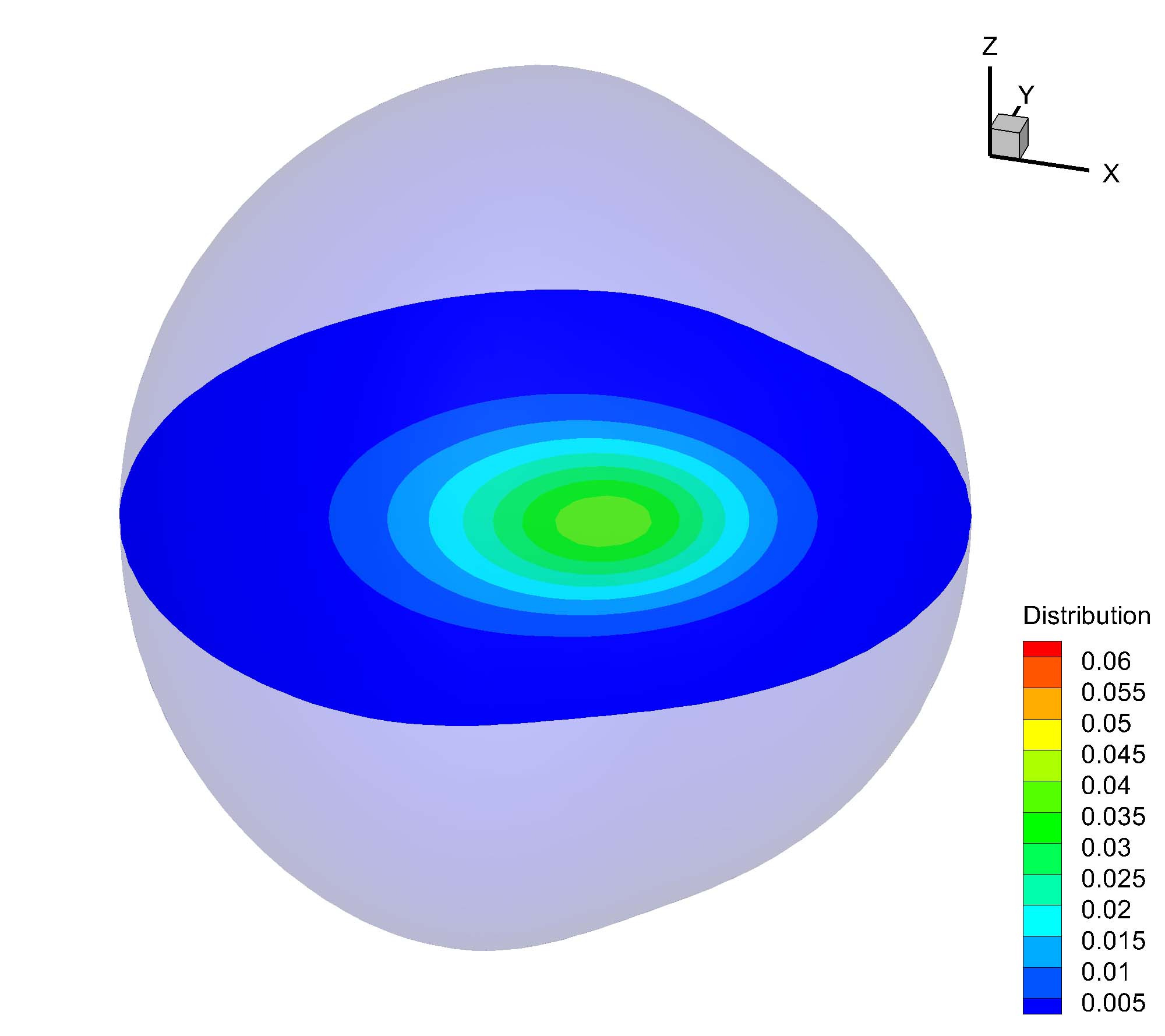}
}
\subfigure[$\tau=2$]{
\includegraphics[width=0.45\textwidth]{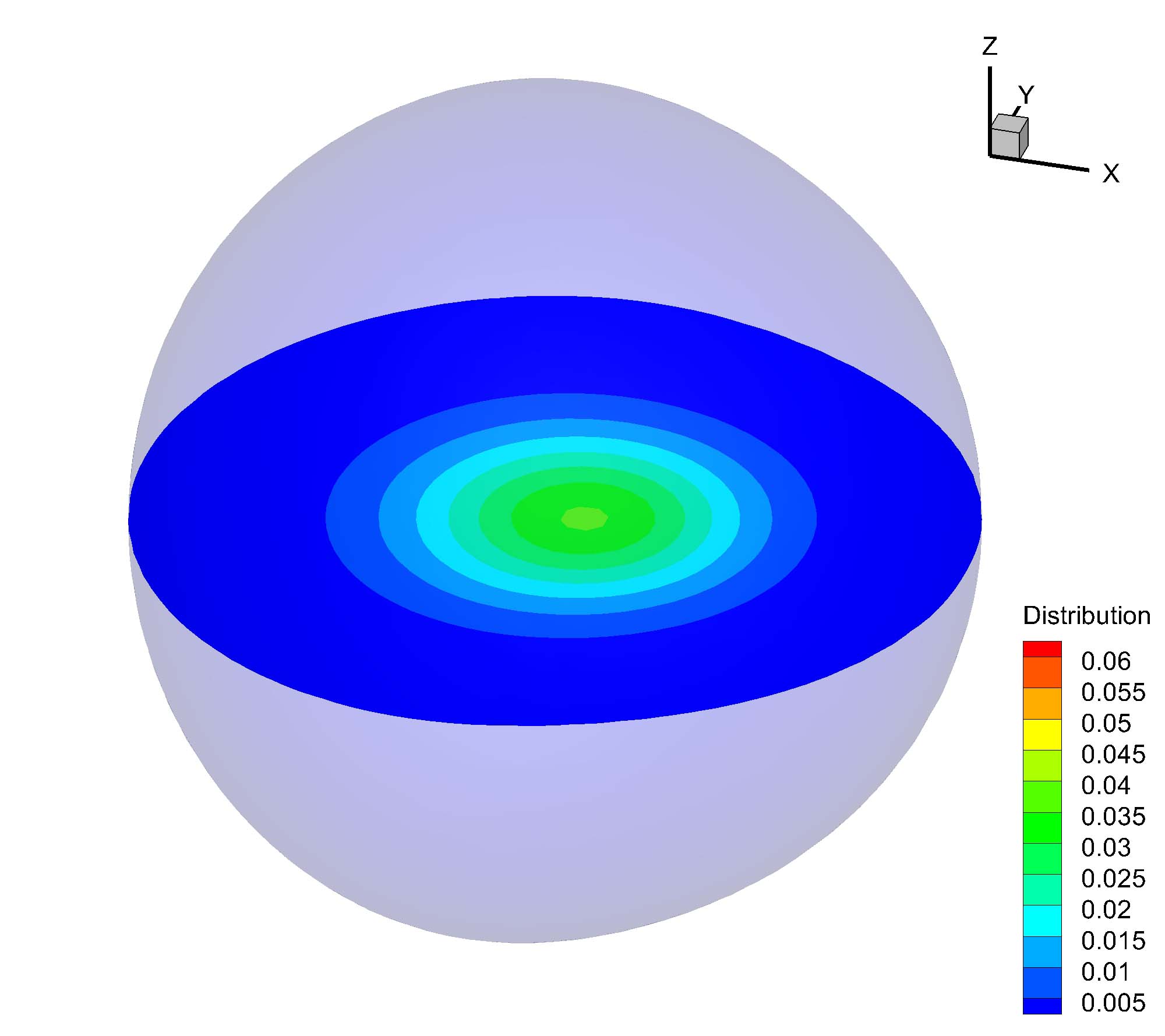}
}
\subfigure[$\tau=10$]{
\includegraphics[width=0.45\textwidth]{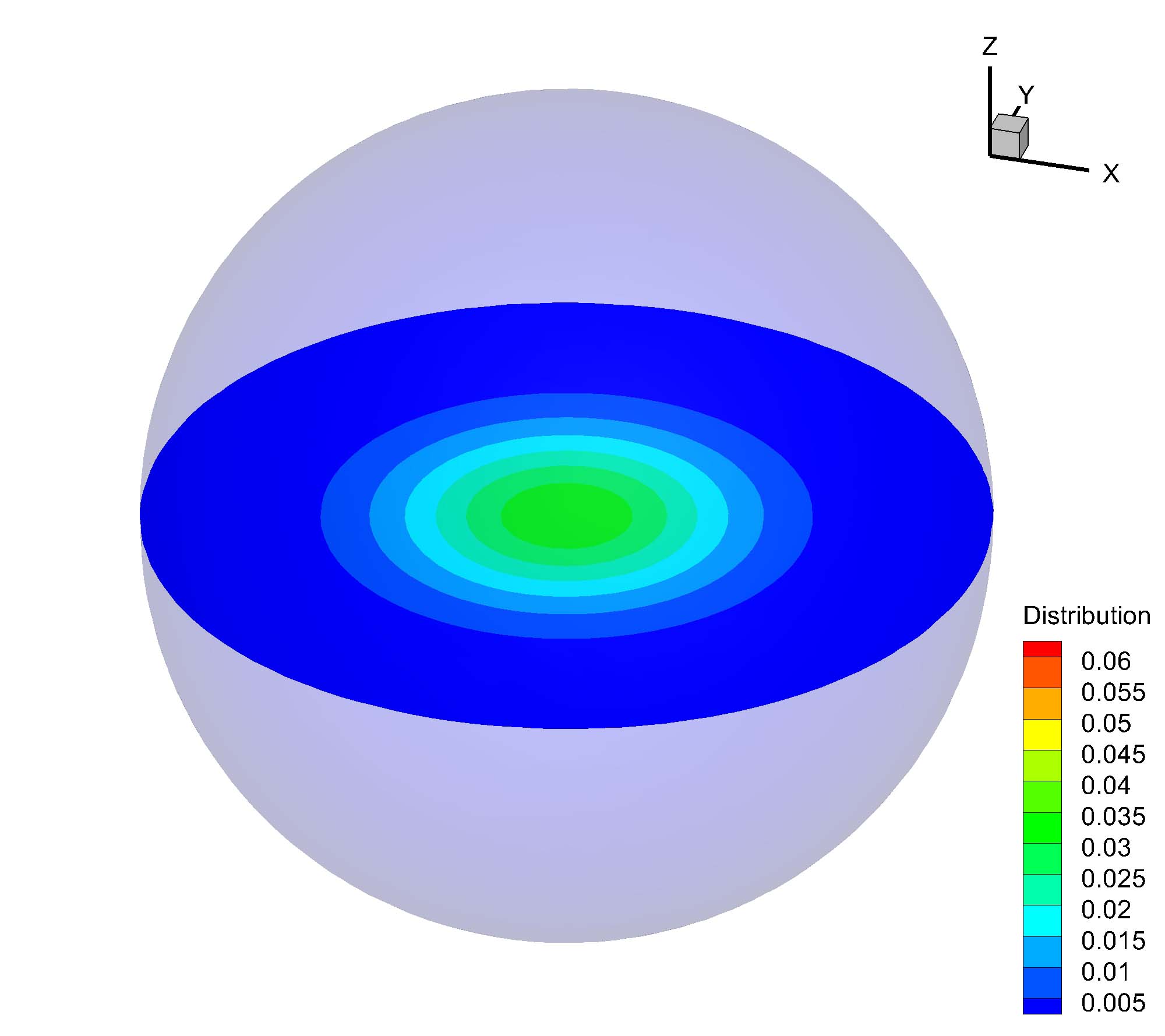}
}
\caption{\label{Fig:discontinu_distribution} The relaxation process of discontinuous distribution, only the region with distribution function greater than $5\times10^{-3}$ is plotted, and the contour is on the plane $\xi_{3}=0$.}
\end{figure}

\begin{figure}
\centering
\subfigure[\label{Fig:case_discontinue_relaxationa} anisotropic temperatures]{
\includegraphics[width=0.45\textwidth]{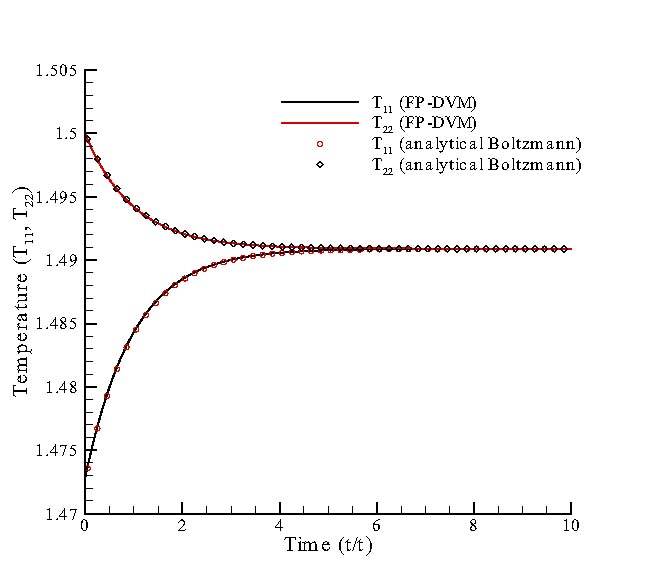}
}\hspace{0.05\textwidth}
\subfigure[\label{Fig:case_discontinue_relaxationb} heat flux]{
\includegraphics[width=0.45\textwidth]{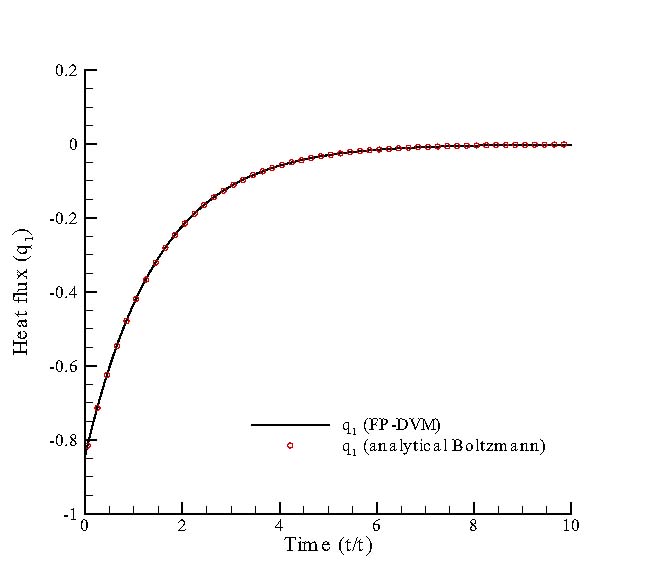}
}
\caption{\label{Fig:case_discontinue_relaxation} The relaxation process of anisotropic temperatures and heat flux of initial discontinuous distribution.}
\end{figure}

The discontinuous distribution function in this case mimics the non-equilibrium distribution at the gas-solid boundary or in Knudsen layer. It is composed of two half Maxwellian distributions. The interface of two half Maxwellian distributions in velocity space is the face $\xi_{1} = 0$. Across the interface, the distribution is discontinuous. The Maxwellian distribution on left is determined from $\rho_a=1.0$, $u_{a,i}=0$, $T_a=2.0$, while the Maxwellian distribution on the right is determined from $\rho_b=1.0$, $u_{b,i}=0$, $T_b=1.0$. This setting mimics the situation that the temperature of fluid is different from the temperature of the solid wall. The truncated discrete velocity space is $[-8,8]$ in each direction with 80 discrete points (80*80*80 mesh). The iteration time $\Delta t$ is chosen as $0.005\tau$. The time evolution of the initial discontinuous distribution is shown in Fig.~\ref{Fig:discontinu_distribution}, where the discontinuity disappears during only one $\tau$, and gradually achieves equilibrium during several $\tau$. The evolutions of anisotropic temperatures and heat flux predicted by the present method are shown in Fig.\ref{Fig:case_discontinue_relaxation}. Due to the exponent term in Eq.~\ref{eq:analytical Boltzmann}, at first the distribution function approaches the equilibrium in a fast rate. This phenomenon can also be seen from the quick disappearance of discontinuity. Then the rate slows down when the distribution is near equilibrium.

\subsection{Normal shock wave structure ($1D$-quasi $1V$ case reduced from $1D$-$3V$ case)}
Shock structure prediction is a benchmark test case for non-equilibrium flow models and corresponding numerical methods. In macroscopic point of view, normal shock wave is a discontinuity in space, across which physical variables change suddenly. While, in microscopic point of view (zoom into the thin shock wave), the physical variables in the shock wave changes smoothly from the front to the back of shock wave. Physically, the molecules in the shock wave are composed of the molecules before the shock (super/hyper-sonic, low temperature) and the molecules after the shock (subsonic, high temperature). When shock $Mach$ number is high, the separation of distribution functions before and after the shock in velocity space is large. Since the molecular collisions in the thin shock wave (about twenty m.f.p.) are insufficient, the distribution function will be far from equilibrium (high non-equilibrium).

Variable Soft Sphere (VSS) model is used in this case since it can be reduced to Hard Sphere (HS) model and inverse power potential model directly by using their scattering factor $\alpha$ and heat index $\omega$. The m.f.p. of VSS model is defined as
\begin{equation}
m.f.p. = \frac{1}{\beta}\sqrt {\frac{{RT}}{{2\pi }}} \frac{\mu }{p},
\end{equation}
where $\beta$ is defined as
\begin{equation}
\beta {\rm{ = }}\frac{{5\left( {\alpha {\rm{ + 1}}} \right)\left( {\alpha {\rm{ + 2}}} \right)}}{{{\rm{4}}\alpha \left( {5 - 2\omega } \right)\left( {7 - 2\omega } \right)}}.
\end{equation}

In shock structure case, the upstream and downstream conditions are determined by Rankine-Hugoniot relation. The iteration time step of FP-DVM is chosen as
\begin{equation}
\Delta t = \min \left( {\Delta {t_{FP}},\Delta {t_{TP}}} \right),
\end{equation}
here the subscript ``TP" stands for ``transport" and corresponds to the free transport operator. $\Delta t_{FP}$  and $\Delta t_{TP}$ can be calculated using the following equation,
\begin{equation}
\begin{aligned}
&\Delta {t_{FP}} = CF{L_{FP}}{\tau _{FP}}\frac{{\Delta \xi }^2}{\max \left( RT \right)},\\
&\Delta {t_{TP}} = CF{L_{TP}}\frac{{\Delta x}}{{\max \left( {{\xi _i}} \right)}}.
\end{aligned}
\end{equation}

\subsubsection{Mach 1.2}
\begin{figure}
\label{Fig:verification}
\centering
\subfigure[\label{Fig:Mach1D2a} density and temperature]{
\includegraphics[width=0.45\textwidth]{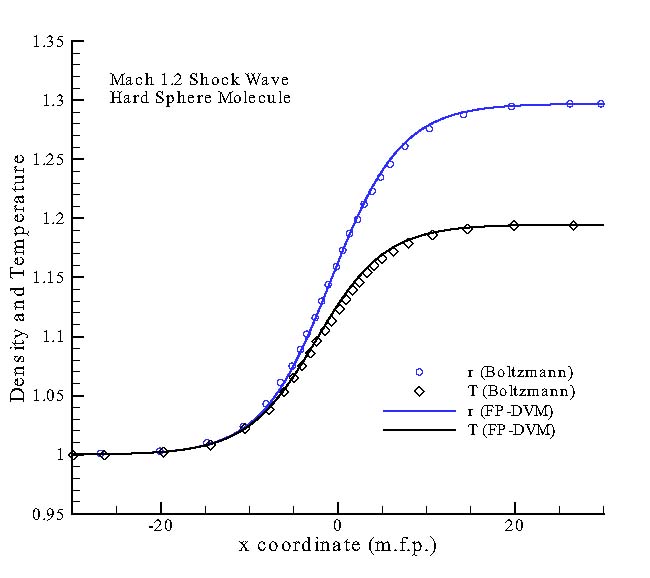}
}\hspace{0.05\textwidth}%
\subfigure[\label{Fig:Mach1D2b} stress and heat flux]{
\includegraphics[width=0.45\textwidth]{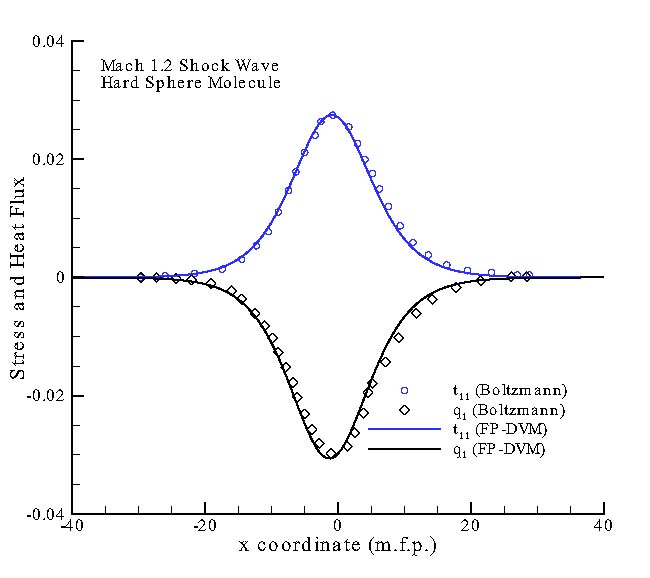}
}
\caption{\label{Fig:Mach1D2} The structures of density, temperature, stress and heat in $Mach=1.2$ shock wave. (full Boltzmann--symbols, FP-DVM--solid lines).}
\end{figure}

In this case, HS molecule model ($\omega = 0.5$, $\alpha = 1.0$) is used which is the same with the deterministic solution of full-Boltzmann equation in Ref.~\cite{Ohwada1993Structure}. For full-Boltzmann solution, x coordinate is non-dimensionalized using the m.f.p. of HS molecule~\cite{Ohwada1993Structure}. The density, temperature, stress, and heat flux in the shock wave are non-dimensionalized using
\begin{equation}\label{eq:nondimensionMach1}
\begin{aligned}
&\hat \rho = \frac{\rho }{{{\rho _{up}}}},~~~\hat T = \frac{T}{{{T_{up}}}},\\
&\hat \tau_{11} = -\frac{\tau_{11} }{{{p_{up}}}},~~~\hat q_{1} = \frac{q_{1}}{{{p_{up}}\sqrt {2R{T_{up}}}}},
\end{aligned}
\end{equation}
here subscripts ``up" and ``down" are used to indicate the variables in the upstream and downstream of shock wave, respectively. The truncated discrete velocity space is $[-7\sqrt{RT_{up}}, 7\sqrt{RT_{up}}]$ with 70 points. The cell Knudsen number $Kn_{cell}={m.f.p.}/{\Delta x}$ is chosen as 4.0 (the cell length in physical space is a quarter of m.f.p.). $CFL_{FP}$ and $CFL_{TP}$ are set 1.0 and 0.9, respectively. The density/temperature profile and stress/heat flux profile are illustrated in Fig.~\ref{Fig:Mach1D2a} and Fig.~\ref{Fig:Mach1D2b}. The FP-DVM predictions match well with the full-Boltzmann result in Ref.~\cite{Ohwada1993Structure}. In this case, the $Mach$ number is low, and the non-equilibrium is not too strong.

\subsubsection{Mach 3.0}
\begin{figure}
\centering
\subfigure[\label{Fig:Mach3a} density and temperature]{
\includegraphics[width=0.45\textwidth]{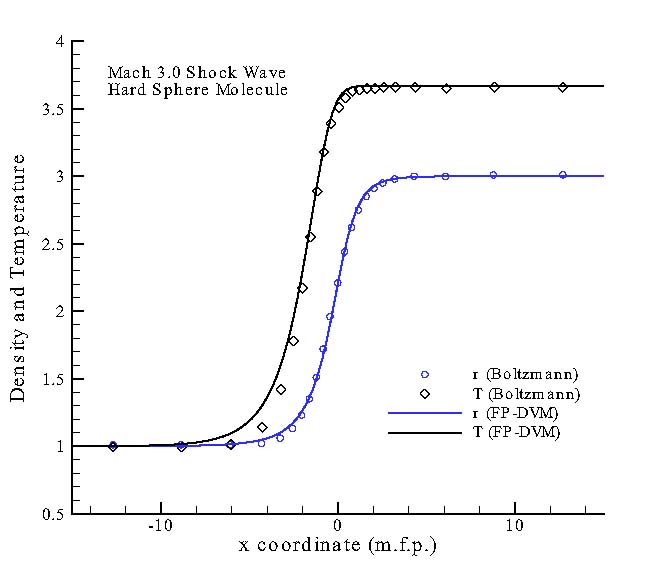}
}\hspace{0.05\textwidth}%
\subfigure[\label{Fig:Mach3b} stress and heat flux]{
\includegraphics[width=0.45\textwidth]{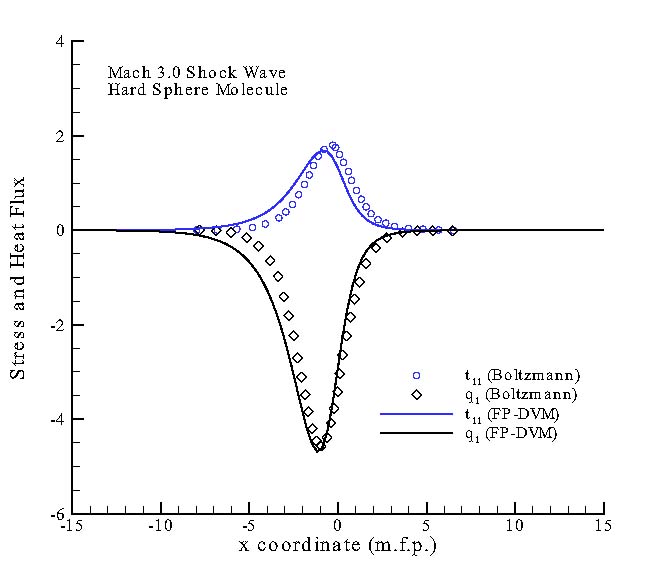}
}
\caption{\label{Fig:Mach3} The structures of density, temperature, stress and heat flux in $Mach=3.0$ shock wave. (full Boltzmann--symbols, FP-DVM--solid lines).}
\end{figure}

The same HS model ($\omega = 0.5$, $\alpha = 1.0$) as $Mach$ 1.2 case is used here along with the same non-dimensionalized x coordinate and physical variable (Eq.~\ref{eq:nondimensionMach1}). Comparing with the $Mach$ $1.2$ case, the degree of non-equilibrium increases in this case. The truncated discrete velocity is $[-10\sqrt{RT_{up}}, 10\sqrt{RT_{up}}]$ with $100$ points. $Kn_{cell}=4.0$ is used. $CFL_{FP}$ and $CFL_{TP}$ are set 1.0 and 0.8, respectively. The density/temperature profile and stress/heat flux profile are illustrated in Fig.~\ref{Fig:Mach3a} and Fig.~\ref{Fig:Mach3b}. The FP-DVM predictions match well with the full-Boltzmann result in Ref.~\cite{Ohwada1993Structure}, expect that the temperature profile rises a little earlier, so are the stress and heat flux profiles.

\subsubsection{Mach 8.0}
\begin{figure}
\centering
\subfigure[\label{Fig:Mach8a} density and temperature]{
\includegraphics[width=0.45\textwidth]{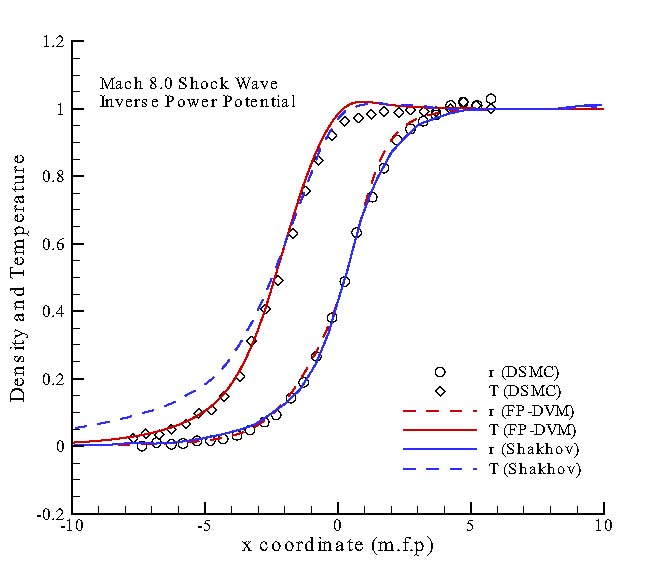}
}\hspace{0.05\textwidth}%
\subfigure[\label{Fig:Mach8b} stress and heat flux]{
\includegraphics[width=0.45\textwidth]{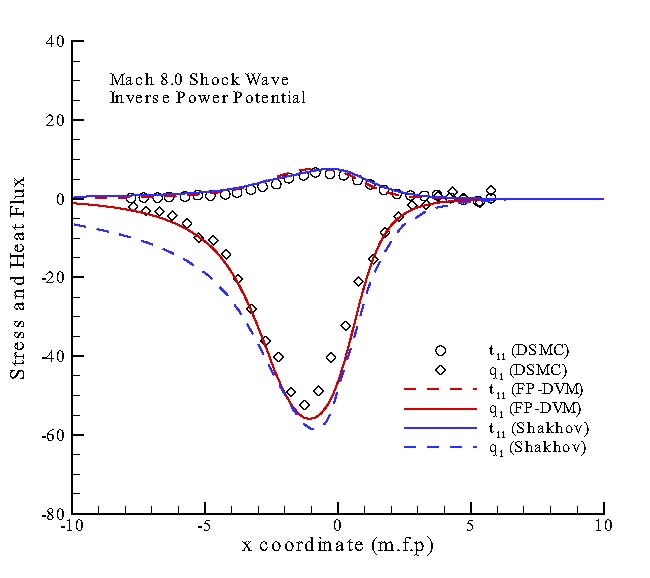}
}
\caption{\label{Fig:Mach8} The structures of density, temperature, stress and heat flux in $Mach=8.0$ Argon shock wave. (DSMC--symbols, FP-DVM--red solid and dash lines, Shakhov--blue solid and dash lines).}
\end{figure}

The working gas is Argon in this case. When $Mach$ number is 8.0, the flow inside the shock wave is in high non-equilibrium. Being the same with Ref.~\cite{Bird1970Aspects}, a 11th power inverse power potential model is used, whose model coefficients can be calculated from Ref.~\cite{Koura1998Variable} as $\omega = 0.68$, $\alpha=1.4225$. The density, temperature, stress and heat flux profiles are calculated and compared with the DSMC results in Ref.~\cite{Bird1970Aspects}. Instead of using the m.f.p. of inverse power potential model and being the same with the setting in Ref.~\cite{Bird1970Aspects}, the x coordinates in the profiles are non-dimensionalized using the m.f.p. of HS molecule. The density and temperature are normalized using
\begin{equation}
\hat \rho {\rm{ = }}\frac{{\rho  - {\rho _{up}}}}{{{\rho _{down}} - {\rho _{up}}}},~~~\hat T = \frac{{T - {T_{up}}}}{{{T_{down}} - {T_{up}}}}.
\end{equation}
The stress and heat flux are non-dimensionalized using
\begin{equation}
\hat \tau_{11} {\rm{ = }}-\frac{\tau_{11} }{{{\rho _{\rm{up}}}{{\left( {2RT _{\rm{up}}} \right)}}}},~~~\hat q_{1} = \frac{q_{1}}{{{\rho _{\rm{up}}}{{\left( {2RT _{\rm{up}}} \right)}^{3/2}}}}.
\end{equation}
The truncated discrete velocity space is $[-30\sqrt{RT_{up}}, 30\sqrt{RT_{up}}]$ with 300 points, and the cell Knudsen number $Kn_{cell}=4.0$. $CFL_{FP}$ and $CFL_{TP}$ are set 1.0 and 0.6, respectively. The density/temperature profile and stress/heat flux profile are illustrated in Fig.~\ref{Fig:Mach8a} and Fig.~\ref{Fig:Mach8b} along with the numerical prediction using BGK-type Shakhov model in Ref.~\cite{xu2011improved}. The temperature, stress and heat flux profiles predicted by Shakhov model deviate from the DSMC results in the front of the shock wave, while the FP-DVM predictions match well with the DSMC results in such a high $Mach$ number and high non-equilibrium case.

\begin{figure}
\centering
\subfigure[\label{Fig:compare_macro} density and temperature]{
\includegraphics[width=0.45\textwidth]{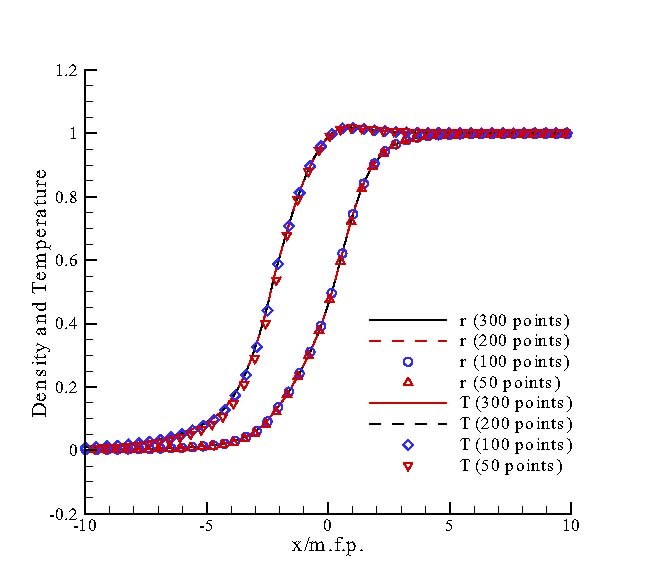}
}\hspace{0.05\textwidth}%
\subfigure[\label{Fig:compare_stress} stress and heat flux]{
\includegraphics[width=0.45\textwidth]{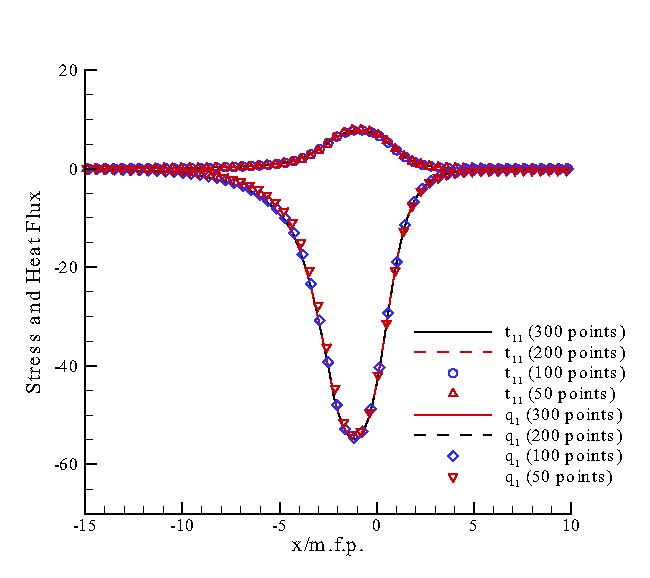}
}
\caption{\label{Fig:compare} The structures of density, temperature, stress and heat flux in $Mach=8.0$ Argon shock wave predicted by FP-DVM with different amount of discrete velocity points in $\xi_{1}$ direction.}
\end{figure}

\begin{figure}
\centering
\subfigure[\label{Fig:F_dis} number distribution F]{
\includegraphics[width=0.45\textwidth]{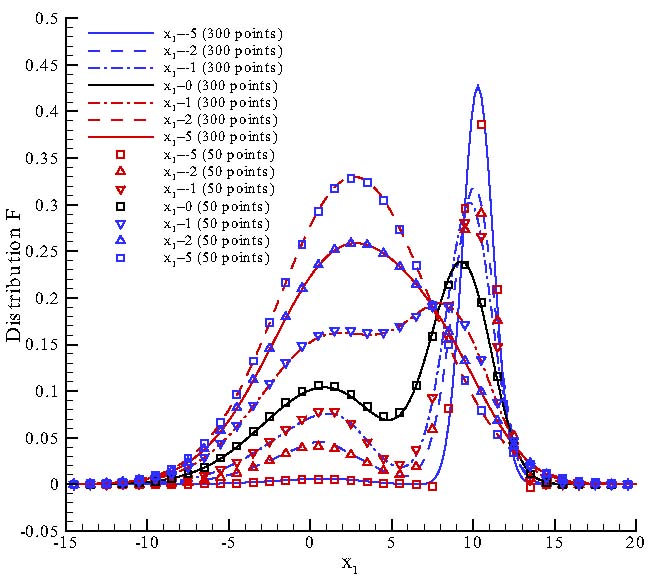}
}\hspace{0.05\textwidth}%
\subfigure[\label{Fig:G_dis} energy distribution G]{
\includegraphics[width=0.45\textwidth]{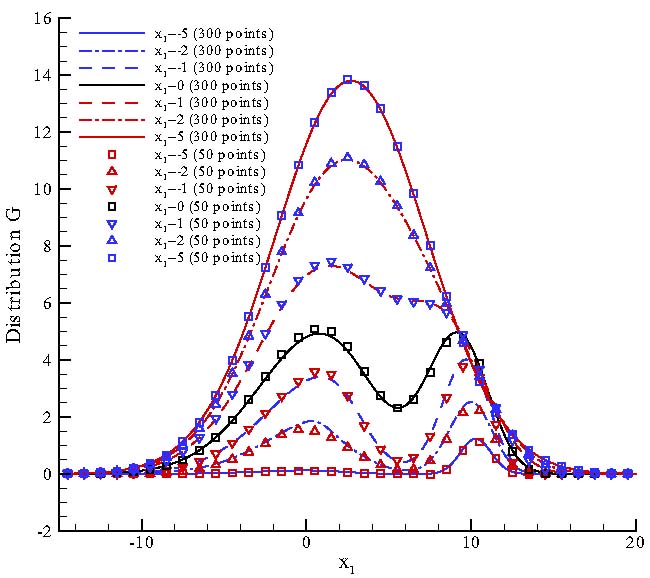}
}
\caption{\label{Fig:FD_dis} The distribution function at different $x_1$ locations inside a $Mach=8.0$ Argon shock wave predicted using 300 and 50 discrete velocity points in $\xi_{1}$ direction, respectively.}
\end{figure}

Since the aim of the above cases is examining the validity of the present FP-DVM method, then the scope of the truncated velocity space and the amount of the discrete velocity points are set to be large. For $Mach$ 8.0 case, the velocity scope $[-25\sqrt{RT_{up}}, 25\sqrt{RT_{up}}]$ is sufficient. With different amount of discrete velocity points in $\xi_{1}$ direction ($300$, $200$, $100$ and $50$ points, respectively), the numerical results predicted by FP-DVM are examined in Fig.~\ref{Fig:compare}. It can be seen that the results with different amount of discrete velocity points almost coincide with each other, expect in the front of the shock wave where the results obtained using $50$ discrete velocity points deviate sightly from the others. The distribution functions $F$ (mass distribution along $\xi_{1}$) and $G$ (energy distribution along $\xi_{1}$) at different locations inside the shock wave are shown in Fig.~\ref{Fig:FD_dis}. They are predicted using $300$ points and $50$ points in $\xi_{1}$ direction, respectively. Since this case is a high non-equilibrium one, the distribution functions inside the shock wave deviate much from the Maxwellian distribution. It can be seen from Fig.~\ref{Fig:FD_dis} that the positive property of distribution function is fulfilled in this high non-equilibrium case. It can also be seen in Fig.~\ref{Fig:FD_dis} that distributions predicted using $50$ discrete velocity points match well with that predicted using $300$ points. Since the discrete velocity space with $50$ points is very coarse, its resolution for precipitous distribution is low. For example, the setting of $50$ discrete velocity points only has $7$ points for approximating the peak of $F$ at $x=-5$ (Fig.~\ref{Fig:F_dis}), and may be the reason of slight deviations in the front of the shock wave (Fig.~\ref{Fig:compare}).

\section{Conclusion}\label{sec:5}
In this paper, a deterministic FP-DVM method is proposed for the non-equilibrium flow simulations. The conservation problem of the discrete ES-FP equation is resolved by multiplying conservative coefficients whose differences with unity are small and have the same orders with the truncation error of the difference scheme for ES-FP collision operator. Using four $0D$-$3V$ cases which mimic different types of distributions that exist in real flow fields, the validity of FP-DVM method and ES-FP model for homogenous cases are proved. In these cases, the evolution of non-equilibrium anisotropic temperatures and heat flux match with the analytical Boltzmann solution precisely. To further extend the scope to ordinary inhomogeneous cases, a $nD$-quasi $nV$ reduction for $nD$-$3V$ ($n<3$) ES-FP equation is developed, which can greatly reduce the computational cost. Using the reduced $1D$-quasi $1V$ FP-DVM method, the shock structure cases from low to high $Mach$ numbers are calculated. The negative distribution function and early rise of temperature profile for high $Mach$ number cases do not appear in the present FP-DVM predictions. All the density, temperature, stress and heat flux profiles match well with the direct full-Botlzmann results and DSMC results. The validity and accuracy of both FP-DVM and ES-FP model for non-equilibrium flow simulation are proved, and the shock structure profiles predicted by the FP-DVM method are probably the best numerical prediction using model Boltzmann equations up to now to the best of our knowledge. Since the FP-type model equations have stiffness problem, the penalty and implicit treatments used in the previous researches should be consider in the further works of the FP-DVM method in order to further increase its iteration time step. The time integral solution of BGK-type equations can also be used in the flux calculation of the present method to extend its scope to whole flow regime (make it a UGKS-type method).

\section*{Acknowledgements}
The authors thank Prof. Kun Xu in Hong Kong University of Science and Technology for discussions of gas kinetic theory and multi-scale flow mechanism. Sha Liu thanks Prof. Jun Zhang in Beihang University and Dr. Fei Fei in Huazhong University of Science and Technology for discussion of Fokker-Planck equation and its numerical method. This work is supported by Fundamental Research Funds for the Central Universities (No. G2018KY0302) and 111 Project of China (No. B17037).

\section*{Reference}
%%\clearpage
\bibliographystyle{elsarticle-num}
\bibliography{Liu_FPUGKS}
\clearpage

\begin{table}
\centering
\caption{\label{table:Maxwell} macroscopic variables and $\varepsilon$ at different settings.}
\begin{tabular}{|c|c|c|c|}
    \hline
    & $50^3$ mesh 2nd order & $50^3$ mesh 4th order & $100^3$ mesh 2nd order\\
    & (base setting)& (high order)& (dense mesh)\\
    \hline
    $\left|\rho-1\right|$ & 8.5e-5 & 8.5e-5 & 1.3e-5\\

    $\left|u_{1}\right|$ & 5.0e-5 & 5.0e-5 & 4.7e-6\\

    $\left|T-1\right|$ & 5.2e-4 & 5.2e-4 & 9.5e-5\\

    $\left|\varepsilon_{f}-1\right|$ & 7.7e-5 & 4.6e-5 & 4.5e-6\\

    $\left|\varepsilon_{A,1}-1\right|$ & 3.0e-5 & 7.4e-5 & 1.1e-6\\

    $\left|\varepsilon_{D}-1\right|$ & 2.0e-2 & 6.4e-4 & 5.0e-3\\

    $\log{\left|\varepsilon_{f}-1\right|}/\log{\Delta \xi}$ & 5.8 & 6.2 & 5.3\\

    $\log{\left|\varepsilon_{A,1}-1\right|}/\log{\Delta \xi}$ & 6.5 & 5.9 & 6.0\\

    $\log{\left|\varepsilon_{D}-1\right|}/\log{\Delta \xi}$ & 2.4 & 4.6 & 2.3\\
    \hline
\end{tabular}
\end{table}

\end{document}